\documentclass[aps,prd,reprint,twocolumn,preprintnumbers,floatfix,nofootinbib]{revtex4-1}
\usepackage[utf8]{inputenc}
\usepackage[dvips]{graphicx}
\usepackage[dvipsnames]{xcolor}
\usepackage{color}
\usepackage{relsize}
\usepackage{graphics}
\usepackage{epstopdf}
\usepackage{hyperref}
\usepackage{mathrsfs}
\usepackage{comment}
\usepackage{ragged2e}
\usepackage{amssymb}
\usepackage{fontawesome} 

\usepackage[normalem]{ ulem }
\usepackage{amsthm}
\usepackage{amsmath}
\usepackage{hyperref}
\usepackage{cancel}
\newcommand{\be}{\begin{equation}}
\newcommand{\ee}{\end{equation}}
\newcommand{\bea}{\begin{eqnarray}}
\newcommand{\eea}{\end{eqnarray}}
\newcommand{\ba}{\begin{aligned}}
\newcommand{\ea}{\end{aligned}}

\newcommand{\dd}{\text{d}}

\newcommand{\gV}{g_{\rm V}}
\newcommand{\gD}{g_{\rm D}}
\newcommand{\mDM}{m_{\rm DM}}

\newcommand{\nDM}{n_{\rm DM}}
\newcommand{\nDMeq}{n_{\rm DM,eq}}
\newcommand{\TRH}{T_{\rm RH}}
\newcommand{\TBH}{T_{\rm BH}}
\newcommand{\MBH}{M_{\rm BH}}

\newcommand{\bh}{{\rm BH}}

\newcommand{\figref}[1]{Fig.~\ref{#1}}

\newcommand{\secref}[1]{Sec.~\ref{#1}}

\usepackage{suffix}
\usepackage{mathtools}

\DeclarePairedDelimiterX\MeijerM[3]{\lparen}{\rparen}%
{\begin{smallmatrix}#1 \\ #2\end{smallmatrix}\delimsize\vert\,#3}

\newcommand\MeijerG[8][]{%
  G^{\,#2,#3}_{#4,#5}\MeijerM[#1]{#6}{#7}{#8}}

\WithSuffix\newcommand\MeijerG*[7]{%
  G^{\,#1,#2}_{#3,#4}\MeijerM*{#5}{#6}{#7}}

\newcommand\myshade{80}
\colorlet{mylinkcolor}{ForestGreen}
\colorlet{mycitecolor}{Red}
\colorlet{myurlcolor}{violet}

\hypersetup{
  linkcolor  = mylinkcolor!\myshade!black,
  citecolor  = mycitecolor!\myshade!black,
  urlcolor   = myurlcolor!\myshade!black,
  colorlinks = true
}

\usepackage{pgfplots}
\newcommand{\equaref}[1]{Eq.~(\ref{#1})}

\pgfplotsset{compat=1.17}

\begin{document}
\sloppy

\preprint{FERMILAB-PUB-21-305-T, NUHEP-TH/21-07, CP3-21-42, IPPP/21/01}

\vspace*{1mm}

\title{Primordial Black Hole Evaporation and Dark Matter Production:\\
II. Interplay with the Freeze-In/Out Mechanism}

\author{Andrew Cheek$^{a}$}
\email{andrew.cheek@uclouvain.be}
\author{Lucien Heurtier$^{b}$}
\email{lucien.heurtier@durham.ac.uk}
\author{Yuber F. Perez-Gonzalez$^{c, d, e}$}
\email{yfperezg@northwestern.edu}
\author{Jessica Turner$^{b}$}
\email{jessica.turner@durham.ac.uk}

\affiliation{$^a$ Centre for Cosmology, Particle Physics and Phenomenology (CP3),
Universit\'e catholique de Louvain, Chemin du Cyclotron 2,
B-1348 Louvain-la-Neuve, Belgium}
\affiliation{$^b$ Institute for Particle Physics Phenomenology, Durham University, South Road, Durham, U.K.}
\affiliation{$^c$ Theoretical Physics Department, Fermi National Accelerator Laboratory, P.O. Box 500,
Batavia, IL 60510, USA}
\affiliation{$^d$ Department of Physics \& Astronomy, Northwestern University, Evanston, IL 60208, USA}
\affiliation{$^e$ Colegio de Física Fundamental e Interdisciplinaria de las Américas (COFI), 254 Norzagaray
street, San Juan, Puerto Rico 00901.}

\begin{abstract}

We study how the evaporation of primordial black holes (PBHs) can affect the production of dark matter (DM) particles through thermal processes. We consider fermionic DM interacting with Standard Model particles via a spin-1 mediator in the context of a Freeze-Out or Freeze-In mechanism. We show that when PBHs evaporate after dominating the Universe's energy density, PBHs act as a source of DM and continuously inject entropy into the visible sector that can affect the thermal production in three qualitatively different ways. We compute the annihilation cross-sections which account for the interactions between and within the PBH produced and thermally produced DM populations, and establish a set of Boltzmann equations which we solve to obtain the correct relic abundance in those different regimes and confront the results with a set of different cosmological constraints. We provide analytic formulae to calculate the relic abundance for the Freeze-Out and Freeze-In mechanism in a PBH dominated early Universe. We identify regions of the parameter space where the PBHs dilute the relic density and thermalization occurs. Furthermore, we have made our code that numerically solves the Boltzmann equations publicly available. \href{https://github.com/earlyuniverse/ulysses}{\faGithub}

\end{abstract}
\maketitle

\section{Introduction}
The observation of the cosmic microwave background (CMB) radiation, and the study of its perturbations, has shed light on the composition of our Universe at the time of recombination \cite{Ade:2015lrj,Ade:2015xua}. As observed by the Planck satellite, the CMB spectrum's inhomogeneities have revealed that our Universe is surprisingly flat and homogeneous. Inflation is a compelling theory that explains this astonishing homogeneity and flatness. It proposes a mechanism to generate cosmological perturbations from the quantum fluctuations of a single scalar field. Furthermore, depending on the specificity of the inflation model considered, such perturbations might have been sufficiently strong that they would have gravitationally collapsed and seeded a population of {\em primordial black holes} (PBHs). This hypothetical population of PBHs would evaporate via Hawking radiation \cite{Hawking:1974rv,Hawking:1974sw} and if their initial mass $M_{i}\lesssim 10^{9}\,\rm{g}$ then they would have evaporated before Big Bang Nucleosynthesis (BBN). 
{Interestingly, such a scenario may lead to the production of dark radiation in the form of gravitons and affect the spectrum of gravitational waves, in a way which could be tested experimentally in the near future \cite{Masina:2020xhk, Arbey:2021ysg, Domenech:2021wkk}. Nevertheless, this possibility remains up to now unconstrained and it is possible that PBHs dominated the early Universe's energy budget.}

In addition to the wealth of data garnered by the CMB spectrum measurements, there is a large body of evidence that dark matter (DM) constitutes approximately 26$\%$ of the Universe's energy budget. All such evidence comes from the gravitational interactions of DM. Nonetheless, hypothesizing interactions of the DM with the Standard Model (SM) provides possible production mechanisms and insights into its nature and origin. The most testable production mechanism is thermal {\em Freeze-Out} (FO)~\cite{STEIGMAN1985375}, as it assumes interactions between dark matter and SM particles were, at some point, frequent enough to couple the two sectors. Alternatively, {\em Freeze-In} (FI)~\cite{McDonald:2001vt,Hall:2009bx} makes no such assumption but requires some small coupling to the SM such that the thermal bath slowly produces the required abundance of DM.

If a population of PBHs existed in the early Universe, they would affect the production of DM. In the scenario that DM is purely gravitationally interacting, and its mass was smaller than the Hawking temperature of the PBHs, PBH evaporation would be the sole source of DM production. Such a scenario has been studied extensively \cite{Matsas:1998zm,Bell:1998jk,Arbey:2021ysg,Khlopov:2004tn,Allahverdi:2017sks,Fujita:2014hha,Lennon:2017tqq,Morrison:2018xla,Masina:2020xhk,Hooper:2019gtx}. Recently it has been shown that purely gravitationally interacting light, fermionic DM ($m_{\rm DM}\lesssim 2\,\rm{MeV}$) would be produced relativistically by PBHs. In this mass regime, redshifting does not sufficiently cool the DM, and it would disrupt small-scale structure formation \cite{Baldes:2020nuv}. The authors of \cite{Auffinger:2020afu} came to a similar conclusion for higher spin DM candidates, and there is considerable tension on the scenario of light DM purely produced from PBH evaporation. In those different analyses, the DM mass is often neglected as compared to the PBH temperature, the geometrical-optics limit is used in order to obtain analytical results{, and the code \textsf{Blackhawk} is used to compute numerically greybody factors}. In Ref.~\cite{paper1}, we have derived both semi-analytically and numerically the phase-space distribution of DM particles of arbitrary spins and masses, using the full greybody factors. We have also studied in thorough detail the effect of the PBH spin on DM production{, enlarging the study of Ref.~\cite{Masina:2021zpu} by solving numerically Boltzmann equations and providing helpful semi-analytical expressions for the number of particles produced during evaporation including the full greybody factors}.

However, the evaporation of PBHs may not account for the production of the whole DM relic abundance. This is the case if the PBH energy fraction is particularly low or if the PBH mass is large enough. In that case, it is necessary to consider other ways to produce DM in the early Universe. In Ref.~\cite{Gondolo:2020uqv, Bernal:2020bjf, Bernal:2020ili}, it was proposed that DM particles could be produced in different manners and, in particular, through the FI and FO mechanisms. It was shown that the two sources of production could conspire non-trivially to produced the DM relic abundance that is measured today and that the mechanism of entropy dilution could play an important role in the case of the FI scenario. 
{A future detection of particle dark matter may tell us one day what is the mass of DM, how strong its interaction with SM particles are, and therefore to which extent the DM relic abundance has been produced through a FI or FO mechanism. The detection of DM could therefore impose  constraints on the density fraction, mass and spin of PBHs where the PBHs  would either overpopulate the DM relic density or dilute it through an entropy injection.} Understanding these implications requires a greater understanding of the possible interplay between interacting DM and PBHs. This work is a step in that direction by improving on the more approximate arguments found in the literature. 

This paper aims to go one step further and study how the presence of evaporating PBHs in the early Universe can affect the dynamics of the FI and FO mechanisms in addition to producing an extra contribution to the relic abundance. We highlight that two major aspects of such an interplay remain, to our knowledge, unexplored in the literature in the context of PBH evaporation:
\begin{enumerate}
    \item A period of PBH domination leads to a non-trivial modification of the Universe's evolution, which cannot simply be reduced to an instantaneous entropy injection and therefore affects both the FI and FO mechanism dynamics in a sizeable region of the parameter space.
    
    \item The existence of a mediator particle $X$ between the dark and the visible sector can lead to an enhancement of the DM interactions, in particular, when $m_{\rm DM}\gtrsim m_X$, either because additional annihilation channels can open up at large energy, or simply because boosted DM particles can accidentally hit $s$-channel resonances while their momentum redshifts with the Universe's expansion.
\end{enumerate}

Whereas the first point was partially addressed in a more general context of early matter domination in the case of the WIMP \cite{Arias:2019uol} and ultraviolet FI \cite{Bernal:2019mhf} it has never been extensively studied, to our knowledge,  either in the case of PBH evaporation or in the case where the FI mechanism operates at an intermediate temperature\footnote{The term {\em intermediate} referring here to the fact that the FI mechanism is neither UV nor IR dominated but takes place on a resonance, at a temperature $T_{\rm RH}> T_{\rm FI}> m_{\rm DM}$.}\footnote{Note that the authors of Ref.~\cite{Bernal:2020bjf} did consider the possibility that the FO takes place during PBH domination but could only treat that case partially since they were not solving Boltzmann's equation.}. The question of the thermalization of the PBH evaporation products was raised in Ref.~\cite{Gondolo:2020uqv} but only in the case of cross-sections scaling like $E^{-2}$. We go beyond such an approximation and include the full energy dependence of the DM model that we consider.

For this purpose, we focus on the simple case where a fermionic DM particle interacts with SM states by exchanging a spin-1 mediator. We derive the appropriate momentum-averaged Boltzmann equations, including PBH evaporation, that are solved numerically.  For this we use the infrastructure of {\tt ULYSSES} \cite{Granelli:2020pim}, a publicly available python package that has been typically used to solve Boltzmann equations associated with leptogenesis. Our code, used throughout this work, is publicly available \footnote{\url{https://github.com/earlyuniverse/ulysses} \href{https://github.com/earlyuniverse/ulysses}{\faGithub}}. The findings of Ref.~\cite{paper1} provide us with full knowledge of the DM particle energy spectrum after they are produced from PBH evaporation. Using this spectrum, we question whether such particles can constitute a sizeable fraction of DM today and explore in which regime they can be expected to thermalize, either with the SM bath or with the pre-existing relic abundance of cold DM particles. 
{We make the simplifying assumption that the population of PBHs have a
monochromatic mass spectrum, which typically is the case when PBHs are produced from inflationary fluctuations or bubble collisions. More complex distributions exist depending on the scenario considered (see e.g. Ref.~\cite{doi:10.1146/annurev-nucl-050520-125911} for a review).}

Our paper is organized as follows: in \secref{sec:model} we present the generic dark matter model we study and its thermal production mechanism. Then, in \secref{sec:PBH},  we summarize the pertinent features of PBHs relevant for this paper and in \secref{sec:interplay} we discuss the interplay between thermal and PBH produced DM as well as how the PBHs can affect the evolution of the Universe which indirectly impacts DM production. The Boltzmann equations we solve, which encapsulate this interplay, are presented in \secref{sec:BEs} and we discuss our approach in order to solve them in the Freeze-In and Freeze-Out cases in \secref{sec:FI} and \secref{sec:FO} respectively. In \secref{sec:analytic},  we derive analytical estimates of the DM relic density produced through Freeze-In and Freeze-Out production in a background modified by the presence of PBHs and in \secref{sec:constraints} we discuss the constraints on the parameter space that we use regarding the presence of warm dark matter.  Finally, in \secref{sec:results} we present the effects of PBH evaporation on the Freeze-Out and Freeze-In mechanism as well as discussing the impact of PBH spin on DM production. 

\section{Thermal Production of dark matter Particles}\label{sec:model}

This Section reviews the generic particle physics model that we will study throughout this work and the two main processes of dark matter production that we will consider, aside from the evaporation of primordial black holes.

\subsection{The Model}
\begin{figure}
 \includegraphics[width=0.9\linewidth]{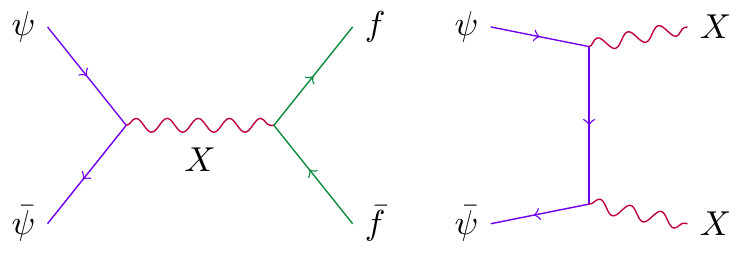}
 \caption{Example of processes leading the production of thermalized DM particles.}\label{fig:diagrams} 
\end{figure}

 We consider the dark matter particle, $\psi$, to be a massive Dirac fermion which is a singlet under the SM gauge group but charged under a dark Abelian symmetry $U(1)_X$. The gauge boson associated with this new symmetry is denoted as $X_\mu$. For simplicity, we assume that the mass of the latter originates from a St\"uckelberg mechanism \cite{Stueckelberg:1900zz}, such that we do not need to assume the existence of any other Beyond Standard Model particles other than the dark matter particle, $\psi$, and the mediator, $X$, throughout this paper\footnote{Note that in principle, the existence of a UV sector may affect the following discussion as heavy particles can be produced towards the end of PBH evaporation. However, because such particles are heavy, they will only be produced in small proportions compared to DM particles and their mediator. Therefore such a contribution to the final relic abundance is expected to be subdominant.}. The Lagrangian can be written
\bea
\mathcal L &=& \mathcal L_{\rm SM}+\bar\psi(i\cancel{\partial}-m_{\rm DM})\psi
+\frac{1}{4}X_{\mu\nu}X^{\mu\nu}-\frac{1}{2}M_X^2 X_\mu X^\mu \nonumber\\ &-& g_{\rm D} X_\mu\bar\psi\gamma^\mu\psi- g_{\rm V} X_\mu\bar f\gamma^\mu f\,,
\eea
where $f$ can be any fermion in equilibrium with the SM bath throughout DM production. For the sake of generality, we assume that $m_{\rm DM}\gg m_f$ and so we can safely neglect SM masses in our analysis. As is illustrated in Fig.~\ref{fig:diagrams}, different processes can lead to the production or thermalization of DM particles from or with the SM bath, including the annihilation of SM fermions into DM particles and the annihilation of vector mediators. 
\subsection{Thermal Production}
Depending on the value of the dark and visible couplings, $\gD$ and $\gV$ respectively, the DM particles may or may not thermalize with the SM bath. In the absence of PBHs, a DM relic density is expected to be produced either through the thermal decoupling of DM particles (the FO scenario), or through their out-of-equilibrium production from the annihilation of SM particles in the plasma (FI scenario). Tracking the evolution of the relic density can therefore be achieved by solving the Boltzmann equation
\be
\dot{n}_{\rm DM} + 3H n_{\rm DM} = -\langle\sigma v \rangle_{\rm th} ( \nDM^2 - \nDMeq^2)\,,
\ee
where $\nDM$ denotes the number density of DM particles $\psi$, $\nDMeq=g_\star \mDM^2 T/(2\pi^2)K_2(\mDM/T)$ is the temperature-dependent number density of DM at thermal equilibrium, and $\langle \sigma v\rangle_{\rm th}$ is the thermally averaged cross-section of DM annihilation. Note that we use the subscript ``th" to emphasize that such a thermal average is performed over the phase-space of SM particles that follow the usual Fermi-Dirac thermal distribution. Although this might sound obvious at this stage, we will see in what follows that keeping track of which thermal distribution one is averaging over will be important when PBHs are introduced. 
Henceforth, we will assume that the mediator mass satisfies  
\be\label{eq:AssumptionMasses}
\TRH > m_X > 2\mDM\,.
\ee
In the case of the Freeze-Out mechanism, this condition ensures that the DM is driven out-of-equilibrium by the $s$-channel annihilation involving the exchange of an off-shell $X$ boson as the $t$-channel annihilation of DM particles into a pair of mediators is kinematically forbidden. While in the Freeze-In case, this guarantees that DM is mainly produced on the resonance when the mediator $X$ goes on-shell at $T\approx m_X$. In that case, we will also assume (and check {\em a posteriori}) that the mediator couples feebly enough to the SM such that it does not reach thermal equilibrium at any time with the SM bath and that the major production mechanism is via 2 $\to$ 2 annihilation processes.

Later we will assume that those two mechanisms of DM production will co-exist with a population of primordial black holes that evaporate before Big-Bang Nucleosynthesis.

\subsection{Non-relativistic parameterization}
In this paper, we will not scan over the values of the dark and visible couplings but will parametrize instead the model in terms of the SM$\Longleftrightarrow$DM annihilation cross-section, $\langle\sigma v\rangle$, and branching fraction of the decay of $X$ into DM particles, $\mathrm{Br}(X\to \mathrm{DM})$. In the non-relativistic limit, the former is given by
\be\label{eq:sigmav}
\langle\sigma v\rangle = \frac{\gV^2\gD^2}{2\pi}\frac{2 \mDM^2+m_f^2}{(4\mDM^2-M_X^2)^2+M_X^2 \Gamma_X^2}\sqrt{1 -\frac{m_f^2}{\mDM^2}}\,,
\ee
where the expression of the mediator decay width, $\Gamma_X$, is given in Appendix~\ref{app:Xsections}.
After it is produced by PBH evaporation, the mediator, $X$, eventually decays back into DM particles, according to its decay branching fraction which, in the limit, $\mDM,m_f\ll m_X$, reads
\be
\mathrm{Br}(X\to \mathrm{DM})\equiv \frac{\Gamma_{X\to \mathrm{DM}}}{\Gamma_X}\sim\frac{\gD^2}{\gD^2+\gV^2}\,.
\ee

When scanning over the parameter space, these relations will allow us, for any value of the non-relativistic cross-section $\sigma v$, and dark branching fraction $\mathrm{Br}(X\to \mathrm{DM})$, to obtain the values of the couplings $\gV$ and $\gD$ as a function of the particle masses $m_X$ and $\mDM$.

\section {Primordial Black Hole Evaporation}
\label{sec:PBH}

As they evaporate, PBHs can produce a population of boosted DM and mediator particles that can contribute to the final relic abundance or thermalize with the SM bath. In Ref.~\cite{paper1} we studied this mechanism of production extensively and described the phase-space distribution of such particles. In this Section, we recapitulate the essential elements and notations that will be used throughout the paper.

The initial PBH mass is taken to scale with the particle horizon mass at the time of PBH formation following~\cite{Carr:2020xqk}
\begin{align}
 \MBH^{\rm in} = \frac{4\pi}{3} \gamma\frac{\rho_{\rm rad}^{\rm in}}{H_{\rm in}^3}\,,
\end{align}
where $\gamma=(1/\sqrt{3})^3\approx 0.2$. The initial fraction of the PBH energy density $\rho^{\rm in}_{\rm PBH}$ when they are formed is related to the radiation energy density $\rho_{\rm rad}^{\rm in}$ via the choice of the parameter $\beta\equiv\rho^{\rm in}_{\rm PBH}/\rho_{\rm rad}^{\rm in} $. However, it is common to express this fraction using the rescaling
\begin{align}
 \beta^\prime \equiv \gamma^{1/2}\left(\frac{g_\star (T_{\rm in})}{106.75}\right)^{-1/4}\beta\,,
\end{align}
where $T_{\rm in}$ denotes {the temperature of the thermal plasma} when PBHs are formed. 

As Hawking showed in~\cite{Hawking:1974rv,Hawking:1974sw}, when the black holes do not have any angular momentum, they typically emit particles with a thermal spectrum whose temperature is inversely proportional to the BH mass:
\begin{align}\label{eq:TBH}
 T_{\rm BH} = \frac{1}{8\pi G M_{\rm BH}}\sim 1.06~{\rm GeV}\left(\frac{10^{13}~{\rm g}}{M_{\rm BH}}\right).
\end{align}
While they evaporate, PBHs lose mass, and therefore, according to \equaref{eq:TBH}, their Hawking temperature increases. Interestingly this implies that the evaporation of PBHs can always produce particles whose masses are as heavy as the Planck mass, in proportions dictated by the evaporation dynamics.
Such dynamics have been extensively studied in the literature and can be described by the following equation~\cite{PhysRevD.41.3052,PhysRevD.44.376}
\bea\label{eq:MBH}
\frac{\dd \MBH}{\dd t}&\equiv &\sum_i \left.\frac{
\dd \MBH}{\dd t}\right|_i = -\sum_i \int_0^\infty E_i \frac{\dd^2 \mathcal{N}_{i}}{\dd p\dd t} dp\nonumber\\ 
&\approx&-\varepsilon (\MBH)\frac{M_p^4}{\MBH^2}\,,
\eea
where $M_p$ denotes the Planck mass. The rate of mass loss will be modified if the PBH has a non-zero angular momentum. Such cases of Kerr PBHs are discussed in detail in our companion paper \cite{paper1}. 
In Eq.~\ref{eq:MBH}, $\dd^2 \mathcal{N}_{i}/\dd p\dd t$ denotes the particle emission rate per PBH for a particle species $i$ of mass $m_i$, spin $s_i$ and number of degrees of freedom $g_i$. This rate can be computed in full generality for any particle spin and mass and was used to derive the spectrum of DM and mediator particles $f_{\rm DM}$ produced through evaporation in Ref.~\cite{paper1}. The so-called mass evaporation function $\varepsilon(\MBH)=\sum_i\varepsilon_i(\MBH)$ are defined in \cite{PhysRevD.44.376,Lunardini:2019zob}. It is particularly useful in order to estimate the amount of energy that is injected by one PBH into each species $i$ at the time of evaporation, which we write as
\bea
\left.\frac{\dd \MBH}{\dd t}\right|_{i}\equiv -\varepsilon_{i} (\MBH)\, \left(\frac{M_p^4}{\MBH^2}\right)\,.
\eea
We refer the reader to Ref.~\cite{paper1} for a detailed study of the emission rates that are used throughout this work.

\begin{figure}[t!]
 \includegraphics[width=0.45\textwidth]{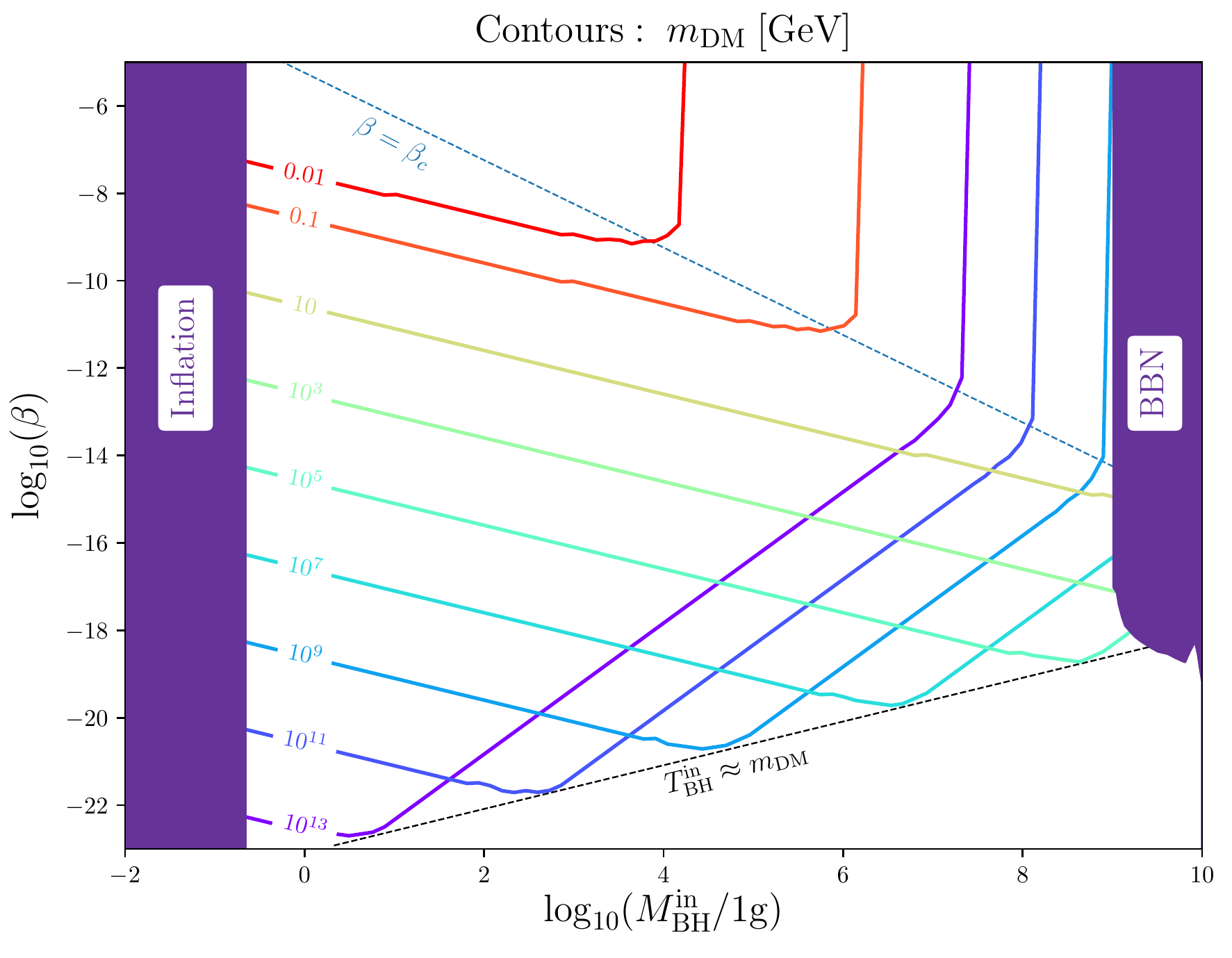}
 \caption{PBH energy fraction $\beta$ as a function of the PBH mass leading to the observed relic abundance $\Omega h^2=0.11$ for different values of the DM mass (in GeV).}\label{fig:DM_PBH}
\end{figure}
Aside from SM states, PBHs emit the mediator and DM particles during their evaporation. In Ref.~\cite{paper1} those modes of production have been studied in depth. In particular, the emission of a spin-1 mediator acts as a secondary source for the production of DM particles, which can be enhanced if the PBHs have a non-zero spin. In Fig.~\ref{fig:DM_PBH} we reproduced the results of Ref.~\cite{paper1} and show the energy fraction and PBH mass that is necessary to obtain the correct relic abundance from PBH evaporation only. For a given choice of the DM mass, the region above the contours leads to an overabundance of dark matter, whereas the region below the contours leads to an underproduction of DM. Another interesting feature of the DM production from PBH evaporation is that in the region where PBHs can dominate the energy density of the Universe before evaporating ($\beta>\beta_c$), the relic density of DM is independent of the energy fraction $\beta$. This is because the energy densities of both the SM and the DM sectors after evaporation are linear in $\beta$. For heavy DM candidates ($m_{\rm DM} \gtrsim 10^9\mathrm{GeV}$) PBHs of large masses need to evaporate significantly before their temperature exceeds the DM mass, which is the case after the relic density contours cross the $\TBH=m_{\rm DM}$ line.

\section{Interplay between PBH Evaporation and thermal production of DM particles}\label{sec:interplay}

Throughout the Universe's evolution, the SM sector, the DM particle, $\psi$, and its mediator, $X$, are assumed to share the universe energy density with a population of primordial black holes. Therefore, the Hubble constant can be written as a function of four elementary contributions
\be\label{eq:Hubble}
3H^2m_p^2=\rho_{\rm SM} + \rho_{\rm DM} + \rho_X + \rho_{\rm PBH}\,,
\ee
where $m_p\equiv M_p/\sqrt{8\pi}$ denotes the reduced Planck mass. Through Hawking evaporation, PBHs are expected to radiate energy and act as a source term for all particle species in proportions dictated by their spin and mass. Consequently, PBHs can efficiently produce many particles, including those associated with a dark sector, regardless of their interaction with or their belonging to the SM sector. 

As we have seen in the previous Section, the evaporation of PBHs constitutes a natural source for the production of DM particles. In addition, DM particles can be produced in our model either from FI or FO mechanism. Although in some instances, those two contributions to the final relic abundance add up without affecting each other, there are different situations in which they interfere, leading to a non-trivial evolution of the relic abundance throughout the Universe history. In Refs.~\cite{Bernal:2020bjf, Bernal:2020ili, Gondolo:2020uqv} some of those cases were studied using a geometrical-optics approach and avoiding solving Boltzmann's equation by focusing on cases that can be studied analytically. In particular, the evaporation of PBHs was treated as an instantaneous process.
Moreover, the findings of such studies strongly depend on the choice of models that was considered: In particular, DM particles were assumed to annihilate into SM states through contact operators (no mediator exchange), and the only Freeze-In scenario considered were either IR or UV dominated. For those reasons, the questions of thermalization of evaporated particles were easily avoided, and the possible evaporation of PBH after FI or FO was reduced to a simple entropy dilution factor in the computation of the relic abundance. In this work, we go one step beyond by studying in detail the temperature dependence of the different DM interactions before and after evaporation. We also treat the evaporation as a continuous process and show that its effect on the cosmological background affects the FI and FO mechanisms non-trivially. We provide a detailed description of those possible effects both analytically and numerically after solving Boltzmann's equations.

\subsection{Modification of the Cosmological Background}\label{sec:modcos}
\begin{figure}
 \centering
 \includegraphics[width=0.95\linewidth]{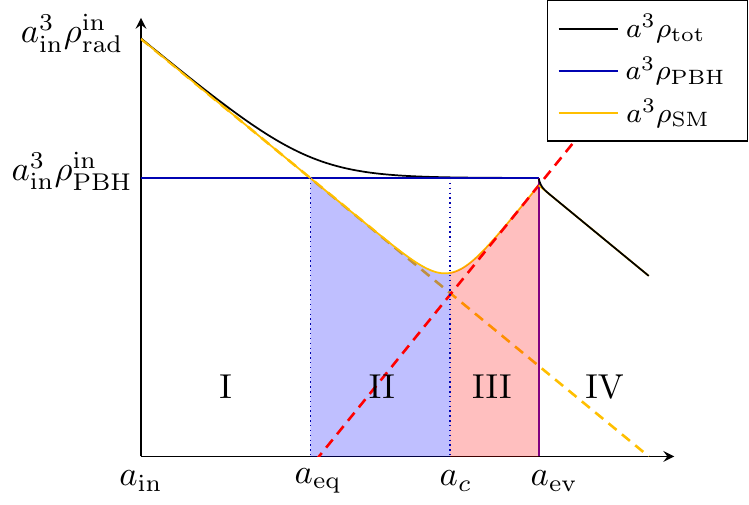}
 \caption{\label{fig:cartoon}\footnotesize Schematic representation of the case where the PBH energy density (blue line) transiently dominates over the SM energy density (orange line). See the main text for a detailed description of the different regimes.}
\end{figure}

Depending on the value of the fraction $\beta$, the presence of PBHs evaporating in the early Universe may or may not modify the evolution of the cosmological background. When $\beta<\beta_c$, PBHs never dominate the energy density of the Universe and their evaporation does not affect its evolution. However, if their energy fraction is such that $\beta >\beta_c$, they can dominate the energy density of the Universe before they evaporate. Because their mass is approximately constant before they evaporate, PBHs behave as a matter component of the Universe with the equation of state parameter $\omega\approx 0$ for $T\lesssim T_{\rm ev}$. In this case, the Universe goes successively from the usual era of radiation domination (regime I in Fig.~\ref{fig:cartoon}, in which $\rho_{\rm tot}\sim \rho_{\rm SM} \propto a^{-4}$) to an early matter-dominated era (regime II) during which the SM bath behaves as radiation ($\rho_{\rm tot}\propto a^{-3}$ and $\rho_{\rm SM}\propto a^{-4}$), followed by a period of entropy injection (regime III) into the SM ($\rho_{\rm tot}\propto a^{-3}$ and $\rho_{\rm SM}\propto a^{-3/2}$). Finally, when evaporation ends, the Universe becomes radiation dominated again (regime IV). 

These four regimes lead to qualitatively different scenarios, depending on when the thermal production of DM particles through FI or FO occurs. There are three scenarios:

\begin{itemize}
 \item [$(i)$] If DM's thermal production occurs during regime I, there is an preexisting relic density of DM particles in the Universe when PBHs evaporate. In that case, the FI or FO mechanism dynamics are not affected by the evaporation, but the remaining relic density gets diluted when the evaporation injects energy into the SM bath. This situation was described in detail in Ref.~\cite{Bernal:2020bjf} in the context of Higgs-portal DM and \cite{Bernal:2020ili} in the context of ultraviolet FI.
  
 \item [$(ii)$] If the thermal production of DM takes place in regimes II and III, the dynamics of the thermal production are modified, and the contribution to the relic density of the FI and FO mechanisms differs from the results derived in a radiation-dominated Universe \cite{Hamdan:2017psw, Arias:2019uol}.
  
 \item[$(iii)$] If thermal processes produce DM in regime IV after PBH have already evaporated, the dynamics of the FI and FO production would, of course, be unaffected by the evaporation of PBHs. The two contributions to the DM relic density may add up together in some instances, but the evaporated  DM particle may also thermalize with the SM bath when produced. 
\end{itemize}

When $\beta>\beta_c$ we have seen that the relic density of DM particles produced through evaporation does not depend on the fraction $\beta$. As one can see in Fig.~\ref{fig:DM_PBH}, DM masses that are sufficiently low ($\mDM\lesssim 1\mathrm{GeV}$) or sufficiently large ($\mDM\gtrsim 10^9\mathrm{GeV}$) do not always lead to an overdensity of DM or violate the BBN bound in the region $\beta>\beta_c$. In those regions, PBHs may dominate the energy density and evaporate later on without overclosing the Universe, while the FI or FO mechanism can occur during any of the phases described above. If PBHs never dominate the Universe's energy density, then the four regimes reduce to one single regime where the Universe evolution is unaffected by the presence of PBHs, and results for the FI and FO production are similar to those obtained in regime IV.

Denoting by $\MBH^{\rm max}(\mDM)$ the value of the PBH mass such that their evaporation produces the correct relic abundance of DM, the condition for the existence of such four regimes can be expressed as
\be
\MBH^{\rm max}(\mDM) < \MBH^{\rm BBN}\,.
\ee
Equivalently, using the results of Ref.~\cite{Gondolo:2020uqv}, this existence bound can be expressed in terms of the temperature of the plasma after evaporation
\be
T_{\rm ev}^{\rm min}(\mDM)>T_{\rm BBN}\,.
\ee

\begin{figure*}[t!]
 \includegraphics[width=\textwidth]{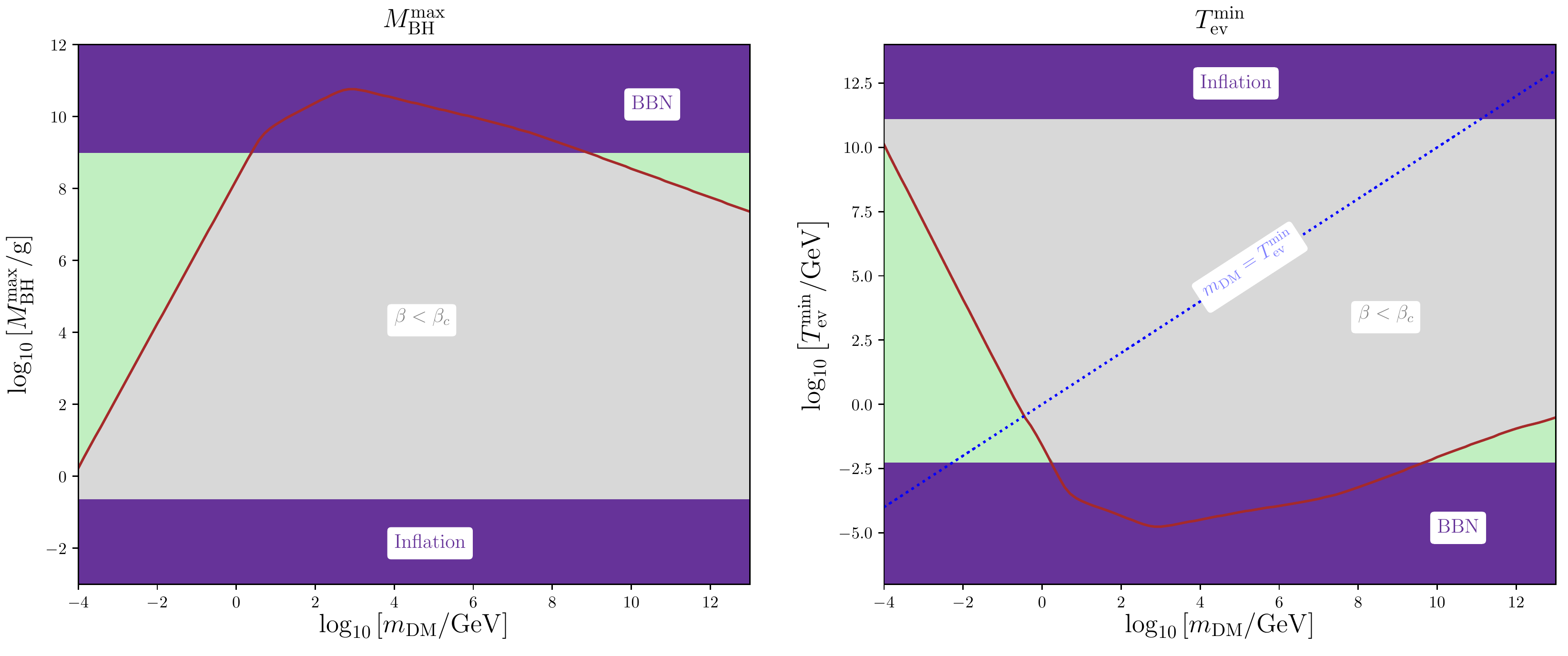}
 \caption{Maximum PBH mass (left panel, brown curve) and the corresponding minimum evaporation temperature (right panel, brown curve) leading to the correct relic abundance of DM particles when it is exclusively produced through PBH evaporation (no mediator, and no BH spin). See the description of the different curves in the text. {The grey-shaded area corresponds to the region where PBHs can only be produced below the critical density in order to not overclose the universe. The green-shaded area corresponds to regions where PBHs are able to dominate the energy density of the universe and evaporate without overclosing the Universe. The blue-dashed line denotes the contour where the PBHs evaporation temperature equals the DM mass.} \label{fig:MBHmin} } 
\end{figure*}
In Fig.~\ref{fig:MBHmin} we show the values of $\MBH^{\rm max}$ and $T_{\rm ev}^{\rm min}$ as a function of the temperature (plain brown line). In the grey-shaded region, PBHs can only evaporate without overclosing the Universe during the radiation-dominated era, which means that the four regimes described above cannot be present in this region of the parameter space. In the green-shaded regions, PBHs can significantly reheat the Universe while producing only a small fraction of the DM relic abundance, opening the possibility that the four regimes exist in that region\footnote{Note that in those regions, PBHs do not necessarily evaporate when dominating the energy density since their energy fraction might as well be subdominant, which is why such regions are not labelled as $\beta<\beta_c$.}. For this reason, if the thermal production of DM particles takes place in the window
\be
T_{\rm ev}^{\rm min}(\mDM)> T_{\rm prod}>T_{\rm BBN}\,,
\ee
the regimes I, II and III may take place (depending on the value of $\beta$), and the FI and FO production mechanisms can be significantly affected by the evaporation of PBHs, as described above.

In the FO mechanism, such a production typically occurs at $T_{\rm prod}\lesssim \mDM$. Therefore, we represented by a blue dotted line the region where $T_{\rm ev}\approx \mDM$ in the right panel of Fig.~\ref{fig:MBHmin}. As one can see, this line only crosses the green-shaded region around DM masses of order $\mDM\sim \mathcal O(10^{-2}-0) \,\mathrm{GeV}$ meaning that the FO mechanism may only be sensitive to the PBH evaporation in this range of DM masses. We discuss this in further detail in \secref{sec:FO}.

However, in the FI mechanism, most of the dark matter production is achieved around $T_{\rm prod}\approx m_X$. Therefore, the dynamics of the FI mechanism may be affected by the evaporation of PBHs only if $T_{\rm ev}\sim T_{\rm prod}\approx m_X$ lies within the green-shaded area. 

It is important to note that those considerations restrict the region of interest in terms of entropy injection {\em during} the DM production mechanism to the low DM mass region. The second green-shaded area, laying over heavier dark matter masses, is a region where the DM relic abundance produced through FI or FO has already been achieved when PBH evaporate. In this region, as it was already described in Refs.~\cite{Bernal:2020bjf, Gondolo:2020uqv}, the entropy injection into the SM bath will lead to a dilution of the preexisting DM relic abundance.

\subsection{Thermalization of Evaporation Products}\label{sec:thermevap}

When they evaporate, PBHs may produce a large population of relativistic DM particles. The interaction rate of such DM particles with SM particles or with DM particles produced through thermal processes strongly depends on the centre-of-mass energy, $E_{\rm com}$, of the processes involved. A large boost factor for evaporated DM particles can open new channels of its annihilation. This is the case when $E_{\rm com}>2 m_X$ since DM particles can annihilate into a pair of mediators through a $t$-channel annihilation ($\bar\psi\psi\to XX$). The $s$-channel annihilation $\bar\psi\psi\to X^\star \to \bar ff$ can also be enhanced as soon as $E_{\rm com}\sim m_X$. Hence, it is crucial to estimate the corresponding interaction rates to consider the possible variation of the DM number density due to such processes when PBHs evaporate.

The thermalization of the evaporated products can have various effects on the DM relic density. In the FI case, because the interaction of DM particles with SM particles is typically much smaller than in the FO case, the thermalization of DM would eventually lead to an overabundance of dark matter and thus be excluded experimentally. In the FO case, the thermalization of evaporated DM particles before the FO has the effect of washing out the contribution of PBHs to the DM relic abundance. If thermalization happens after DM particles have decoupled from the SM bath, the evaporation of PBHs may destroy the predictions of the FO mechanism or recreate initial conditions for a new FO mechanism to take place. In that case, a careful treatment of the Boltzmann equation is required to track the evolution of the DM phase-space distribution after evaporation and the evolution of the relic density. We leave such calculations for future work but nevertheless estimate in the regions of parameter space where this level of calculation will be required.

\section{Boltzmann Equations}\label{sec:BEs}
 This Section details the Boltzmann equations that we will use to track the evolution of the DM relic abundance. In addition, we include terms that track the exchange of particles between the visible and the dark sectors and terms that account for secondary production sourced by the evaporation of primordial black holes.

As mentioned above, the evaporation of PBHs transforms mass and rotational energy into a particle number. Therefore PBH evaporation acts as a source for the phase-space density distribution of the different particle species. This source term has to be defined such that energy is conserved during the evaporation process. Denoting the number density of PBHs of mass $\MBH$ evaporating in the Universe by $n_\bh$, let us define the contribution to the phase-space density distribution time derivative of the species $i$ as 
\be
\left.\frac{p^2}{2\pi^2}\frac{\partial f_i}{\partial t}\right|_\bh(t,p)=\frac{n_\bh}{g_i}\frac{\dd^2 \mathcal{N}_{i}}{\dd p\dd t}\,,
\ee
where $g_i$ is the number of degrees of freedom of the species $i$.
With such a definition, the amount of energy created in the form of particles by PBH evaporation is
\be
\frac{\dd \rho_{\rm ev}}{\dd t}=\sum_i \int_0^\infty E_i \left.\frac{\partial f_i}{\partial t}\right|_\bh\frac{p^2\dd p}{2\pi^2}=-n_\bh \frac{\dd M}{\dd t}\,,
\ee
which exactly compensates the mass-energy loss in the PBH sector. At the level of the phase-space density distribution for the species $i$, it is therefore consistent to write a Boltzmann equation incorporating such a transfer of energy:
\be
\frac{\partial f_i}{\partial t}-Hp\frac{\partial f_i}{\partial p}=C[f_i]+\left.\frac{\partial f_{i}}{\partial t}\right|_\bh\,,
\ee
in which the collision kernel $C[f_i]$ contains the different number-changing interaction rates involving the species $i$. Integrating over the phase-space, one obtains the Boltzmann equation in terms of number densities
\be
\dot{n}_{i} + 3H n_{i} = g_i \int C[f_i]\frac{\dd^3 p}{(2\pi)^3} +\left.\frac{\dd n_{i}}{\dd t}\right|_{\bh}\,,
\ee
where we have defined
\be
\left.\frac{\dd n_i}{\dd t}\right|_\bh=n_\bh g_i\int\left.\frac{\partial f_{i}}{\partial t}\right|_\bh\frac{p^2\dd p}{2\pi^2}\,.
\ee
Together with Eq.~\eqref{eq:Hubble} and Eq.~\eqref{eq:MBH} we can write
\begin{subequations}\label{eq:BoltzDM}
\bea\
\dot{n}_{\rm DM} + 3H n_{\rm DM} &=& g_{\rm DM} \int C[f_{\rm DM}]\frac{\dd^3 p}{(2\pi)^3} +\left.\frac{\dd n_{\rm DM}}{\dd t}\right|_{\bh}\,,\notag\\ \\
\dot{n}_{X} + 3H n_{X} &=& g_{X} \int C[f_{X}]\frac{\dd^3 p}{(2\pi)^3} +\left.\frac{\dd n_{X}}{\dd t}\right|_{\bh}\,,\\
\dot{\rho}_{\rm SM} + 4H \rho_{\rm SM} &=& \left.\frac{\dd M}{\dd t}\right|_{\rm SM}\,.
\eea
\end{subequations}
The difficulty of numerically solving such equations consists of tracking the time evolution of the complete phase-space density distributions of the different species such that the collision term can be evaluated accurately at each time step. Doing so requires considerable computational resources in order to scan over the parameter space. We will thus not attempt to solve the full Boltzmann equations in this current paper. Instead, we will solve the momentum-averaged 
Boltzmann equations that track the number density of dark matter as a function of time. We estimate 
when such an approach is valid and work in regimes where either:
$(i)$ the contributions of the evaporation to the different number densities can be secluded from the particle production from the plasma (which is typically the case in the FI scenario), such that their evolution can be traced independently, or $(ii)$ the particles produced from evaporation quickly thermalize with the plasma (which is typically the case in the FO scenario when the evaporation takes place \emph{before} the dark matter freezes out). In the final Section, we will question the validity of such approximations and probe the parameter space regions in which a better treatment of the phase-space distribution evolution has to be employed. Let us now establish the Boltzmann equations that we are to use to describe the FI and FO cases.

\subsection{Freeze-In Case}
\label{sec:FI}
In the Freeze-In scenario, the DM particles are only very feebly coupled to the SM bath. For that reason, it is reasonable to assume --- and we will check the validity of that assumption in the next section --- that once DM is produced either through thermal processes or evaporation, neither the DM nor mediator thermalizes at any time throughout the universe history. Such a regime can be easily obtained by taking the limit $\mathrm{Br}(X\to \mathrm{SM})\to 0$ (see e.g. Ref.~\cite{Cosme:2021xjf} for a recent review). The calculation of the final dark matter relic density corresponds to summing up the three main contributions
\begin{itemize}
 \item [$(i)$] PBH $\to$ DM\,,
 \item [$(ii)$] PBH $\to$ $X$ $\to$ DM+DM\,,
 \item [$(iii)$] SM+SM $\to$ DM+DM\,,
\end{itemize}
while tracking the evolution of the SM energy density in order to take into account any effects related to the injection of entropy from PBH evaporation as described in the previous section. Because we work in a regime where the particles produced from evaporation --- processes $(i)$ and $(ii)$ --- never thermalize, neither with SM particles, nor within the dark sector, it is convenient to track the separate evolutions of the DM abundance produced from evaporation (processes $(i)$ and $(ii)$) and the abundance produced from thermal processes (process $(iii)$). Denoting the corresponding number densities by $n_{\rm DM}^{\rm ev}$ and $n_{\rm DM}^{\rm th}$, we can rewrite Eq.~\eqref{eq:BoltzDM} as 
\begin{subequations}\label{eq:BoltzDMFI}
\bea
\dot{n}_{\rm DM}^{\rm ev} + 3H n_{\rm DM}^{\rm ev} &=& \left.\frac{\dd n_{\rm DM}^{\rm ev}}{\dd t}\right|_{\bh}+2\, \Gamma_{X\to {\rm DM}}\left\langle \frac{m_X}{E_X}\right\rangle_{\rm ev} n_X\,,\notag\\ \\
\dot{n}_{\rm DM}^{\rm th} + 3H n_{\rm DM}^{\rm th} &=& \left\langle\sigma v\right\rangle_{\rm th}(n_{\rm DM,eq}^2-n_{\rm DM}^{{\rm th}\, 2})\,,\\
\dot{n}_{X} + 3H n_{X} &=& - \Gamma_{X}\left\langle \frac{m_X}{E_X}\right\rangle_{\rm ev}n_X\,,\\
\dot{\rho}_{\rm SM} + 4H \rho_{\rm SM} &=& \left.\frac{\dd M}{\dd t}\right|_{\rm SM}+2 m_X \Gamma_{X\to{\rm SM}}n_X\,.
\eea
\end{subequations}
Note that in the above equations, the distinction between the thermal average over the Boltzmann distribution of SM particles in the plasma denoted by $\langle.\rangle_{\rm th}$ and the average over the phase-space distribution of the mediator X denoted by $\langle.\rangle_{\rm ev}$ is crucial since the mediator particles produced from evaporation can have an average energy $\langle E_X\rangle_{\rm ev}\gg T$.

Denoting the phase-space density distribution of the mediator particles produced through evaporation as $f_{\rm ev}$, the averaged decay width used in Eq.~\eqref{eq:BoltzDMFI} can be expressed as
\be
\Gamma_{X}\left\langle \frac{m_X}{E_X}\right\rangle_{\rm ev}\equiv \Gamma_{X}\int \frac{m_X}{E_X}f_{\rm ev}(p_X)\frac{\dd^3p_X}{(2\pi)^3}\,.
\ee
When discussing the possible thermalization of the evaporation products, such a distinction between the momentum averaged performed over the thermal or evaporated distribution will also be of great importance, as we will see in the next Section.

\subsection{Freeze-Out Case}

\label{sec:FO}
In the FO case, the interactions between the visible and the dark sectors are strong enough to establish thermal equilibrium between the two sectors in the early Universe. After the temperature decreases, the small number density of DM particles together with a decrease of the plasma temperature enforces the decoupling of DM particles from the thermal bath. Similarly to the Freeze-In case, the PBH evaporation acts as a secondary source of DM particles. However, in the FO case, the latter particles may or may not thermalize with the SM bath when they are produced. As it was discussed in the previous Section, we will only consider the two simple cases described below:\newline

\paragraph{Evaporation Before FO with Thermalization}~\newline

If the evaporation occurs before the FO of DM particles from the plasma, we only consider the regime in which the DM particles produced through evaporation instantaneously thermalize with the SM bath. In that case, the contribution of the PBHs to the relic density is washed out, and after evaporation, the normal FO mechanism takes place. Because thermal processes between DM and SM particles are active during evaporation, it is straightforward to realize that the mediator is also thermalized at the time of evaporation. In that case, there is no need to treat the evolution of the DM and mediator number density produced through evaporation separately, and the Boltzmann equations reduce to the usual FO Boltzmann equations with a source term for radiation 
\begin{subequations}\label{eq:BoltzDM_FO_I}
\bea
\dot{n}_{\rm DM} + 3H n_{\rm DM} &=& \left\langle\sigma v\right\rangle_{\rm th}(n_{\rm DM,eq}^2-n_{\rm DM}^2)\,,\\
\dot{\rho}_{\rm SM} + 4H \rho_{\rm SM} &=& \frac{\dd M}{\dd t}\,.
\eea
\end{subequations}

\paragraph{Evaporation After FO without Thermalization}~\newline

If the FO mechanism occurs before the evaporation, the DM particles produced from evaporation may constitute a significant fraction of the DM relative abundance at the time of evaporation as those particles are expected to be significantly boosted. If those DM particles interact efficiently with the DM particles produced earlier via FO, it may significantly alter the FO results and lead to a non-trivial evolution of the DM relative abundance later. In order to avoid such complication, we will evaluate the capacity of the evaporation products to interact with thermally produced particles after evaporation and ensure that we consider only the case where the DM particles produced from evaporation are not able to interact efficiently after they are produced (see details in \secref{sec:constraints}). In that case, we do not need to solve the full Boltzmann equations for the DM and mediator phase-space density distributions, and the different contributions to the relic density add up, similarly to the FI case. Therefore, the Boltzmann equation to consider is the same as in Eq.~\eqref{eq:BoltzDMFI}.

\section{Analytical Treatment}\label{sec:analytic}
As we have seen in \secref{sec:interplay}, the evaporation of PBHs whose energy fraction verifies $\beta>\beta_c$ leads to a modification of the cosmological background evolution. In this Section, we derive analytical estimations of the DM relic density produced through the Freeze-In and Freeze-Out mechanisms in the different regimes that we have exhibited in \secref{sec:interplay}.  The results regarding the Freeze-Out can be found in Ref.~\cite{Arias:2019uol} in a more general context as we have adapted them to the case of PBH evaporation. It is important to note that the following expressions stand for the relic density produced {\em exclusively} through thermal processes and not from evaporation. If PBHs would produce a sizeable contribution to the DM relic abundance when they evaporate, their contribution will have to be added to our present results. An analytic derivation of the DM relic abundance produced solely from PBH evaporation, consistent with our numerical calculations up to an $\mathcal{O}(1)$ factor with that  given in \cite{paper1}.

\subsection{Cosmological Background Evolution}
Before we proceed with the calculation of the relic density, it is necessary to determine the values of the scale factor at the times when PBHs start dominating the energy density ($a_{\rm eq}$), at the critical time when energy injection starts modifying the behavior of the SM bath ($a_c$) and at evaporation ($a_{\rm ev}$). The evolution of the PBH and SM energy densities will then be obtained using appropriate scaling relations between those reference points. At first, SM radiation and PBHs can be described by the scaling relations
\bea
\rho_{\rm PBH}^{\rm I/II/III}(a)&=&\rho_{\rm PBH}^{\rm in}\left(\frac{a_{\rm in}}{a}\right)^3\,,\nonumber\\
\rho_{\rm rad}^{\rm I/II}(a)&=&\rho_{\rm rad}^{\rm in}\left(\frac{a_{\rm in}}{a}\right)^4\,,
\eea
where $\rho_{\rm PBH}^{\rm in}=\beta \rho_{\rm rad}^{\rm in}$.
As we will see, the scaling relation of the radiation bath in regime III differs from one of regimes I and II as the entropy injection into the SM becomes relevant at this point. At matter-radiation equality, one has
\be
a_{\rm eq}=a_{\rm in}/ \beta\,.
\ee
According to Ref.~\cite{Perez-Gonzalez:2020vnz}, the energy loss of PBHs is given by
\be
\frac{\dd \MBH}{\dd t}=-\varepsilon(\MBH) \frac{M_p^4}{\MBH^2}\,,
\ee
which allows us to write
\be
\dot \rho_{\rm PBH} + 3\rho_{\rm PBH}=-\Gamma_{\rm PBH}\times \rho_{\rm PBH}\,,
\ee
where the effective decay constant 
\be\Gamma_{\rm PBH}\equiv \varepsilon(\MBH) \frac{M_p^4}{\MBH^3} \approx \varepsilon(\MBH^{\rm in}) \frac{M_p^4}{\MBH^{\rm in\ 3}}\,,
\ee
is approximately constant during most of the evaporation process. For that reason, PBHs nearly behave like matter ($w\approx 0$) during most of the evaporation phase, and the SM energy density scales like $\propto a^{-3/2}$ during that period of time~\cite{Arias:2019uol}. Using that approximation, PBHs are expected to evaporate when $\Gamma_{\rm PBH}\equiv \nu H_{\rm ev}$ where $\nu$ is an $\mathcal O(1)$ parameter that has to be conveniently chosen. During PBH domination, this occurs when
\be
a_{\rm ev}=a_{\rm in}\left(\frac{8 \pi \nu^2 \rho_{\rm PBH}^{\rm in}\MBH^{\rm in~6}}{3 M_p^{10} \varepsilon(\MBH^{\rm in})^2 }\right)^{1/3}\,,
\ee
where $m_p$ stands for the reduced Planck mass. From this estimation, one obtains the evaporation temperature by assuming instantaneous thermalization of the evaporation products, and we have
\be\label{eq:ev}
T_{\rm ev}\approx\left(\frac{90}{g_{\star,{\rm ev}}\nu^2\pi^2}\frac{\varepsilon_{\rm SM}}{\varepsilon}\Gamma_{\rm PBH}^2 m_p^2\right)^{1/4}\,.
\ee
During the phase of significant entropy injection into the SM (regime III), the energy density of the SM bath can be written
\be
\rho_{\rm rad}^{\rm III}(a)=\rho_{\rm PBH}^{\rm ev}\left(\frac{\varepsilon_{\rm SM}}{\varepsilon}\right)\left(\frac{a_{\rm ev}}{a}\right)^{3/2}\,.
\ee
The critical temperature at which the regime III starts can therefore be expressed by demanding that $\rho_{\rm rad}^{\rm III}(a_c)\equiv \rho_{\rm rad}(a_c)$ giving
\be
a_c=a_{\rm ev}\left(\beta\frac{a_{\rm ev}}{a_{\rm in}}\frac{\varepsilon_{\rm SM}}{\varepsilon}\right)^{-2/5}\,.
\ee
The temperatures $T_{\rm eq}$ and $T_c$ can easily be obtained by using the entropy conservation in the SM bath. 

\subsection{Freeze-In Mechanism}

In order to estimate the amount of DM produced through the Freeze-In mechanism, we solve the following Boltzmann equation
\bea
&&\frac{\dd}{\dd t}\left(a^3 n_{\rm DM}\right)=\nonumber\\
&&\frac{a^3T}{512\pi^5}\int_{4m_{\rm DM}^2}^\infty |\mathcal M|^2\sqrt{s-4m_{\rm DM}^2}K_1\left(\frac{\sqrt{s}}{T}\right)\dd s\,.\nonumber\\
\eea
Because most of the DM production takes place on the resonance, and because couplings of the mediator to DM and SM particles are typically small in the Freeze-In mechanism, it is reasonable to use the narrow-width approximation in order to compute the integral on the center of mass energy $\sqrt{s}$. In the limit where $m_f\to 0$, and rewriting the time derivative in terms of the scale factor, we obtain
\be\label{eq:boltzFI}
\frac{\dd}{\dd a}\left(a^3 n_{\rm DM}\right)=\alpha\, m_X^3 \frac{a^2}{H}T K_1\left(\frac{m_X}{T}\right)\,,\nonumber\\
\ee
where
\bea
\alpha &\equiv&\frac{3 g_D^2 g_V^2 g_{\rm DM}^2}{512\pi^4}
\left(1+2\frac{m_{\rm DM}^2}{m_X^2}\right)\sqrt{1 - 4 \frac{m_{\rm DM}^2}{m_X^2}}\left(\frac{m_X}{\Gamma_X}\right)\,.\nonumber\\
\eea
Solving the above equation in the four different regimes exhibited in Fig.~\ref{fig:cartoon} thus allows us to express $H$ and $T$ as a function of the scale factor in each regime and integrating from $T\gg m_X$ to $T\ll m_X$. The following scaling relations are used in what follows for the calculation of the relic density:
\bea
\text{Regime } & \text{I:}\quad & H^2\propto a^{-4}\quad \text{ and }\quad T\propto a^{-1}\,,\nonumber\\
\text{Regime } & \text{II:} \quad & H^2\propto a^{-3}\quad \text{ and }\quad T\propto a^{-1}\,,\nonumber\\
\text{Regime } & \text{III:}\quad & H^2\propto a^{-3}\quad \text{ and }\quad T\propto a^{-3/8}\,,\nonumber\\
\text{Regime } & \text{IV:}\quad & H^2\propto a^{-4}\quad \text{ and }\quad T\propto a^{-1}\,.\nonumber\\
\eea
Integrating Eq.~\eqref{eq:boltzFI}, we obtain for those different regimes the following estimations:
\begin{widetext}
\begin{subequations}
\bea
\Omega_{\rm I} &=&\alpha\, m_X^3\frac{m_{\rm DM}}{\rho_c}\frac{36\sqrt{10}}{\pi\sqrt{g_{\star,\rho}(m_X)}}\frac{g_{\star,s}(T_{\rm eq})}{g_{\star,s}(m_X)} \frac{T_{\rm eq}^3 m_p}{m_X^4}\frac{a_{\rm eq}^3 }{a_0^3 }\MeijerG*{2}{1}{1}{3}{1}{\frac{3}{2}, \frac{5}{2}, 0}{\frac{m_X}{T_{\rm eq}},\frac{1}{2}}\,,\\
\Omega_{\rm II}&=&\frac{\alpha\, m_X^3}{4}\frac{m_{\rm DM}}{\rho_c}\sqrt{\frac{3m_p^2}{\rho_{\rm PBH}^c}} \left(\frac{a_c}{a_0}\right)^3T_c\left(\frac{g_{\star,s}(T_c)}{g_{\star,s}(m_X)}\right)^\frac{1}{3}\MeijerG*{2}{1}{1}{3}{-\frac{3}{4}}{-\frac{1}{2}, \frac{1}{2}, -\frac{7}{4}}{\frac{m_X}{2T_c}\left(\frac{g_{\star,s}(m_X)}{g_{\star,s}(T_c)}\right)^\frac{1}{3},\frac{1}{2}}\,,\\
\Omega_{\rm III}&=&2 \alpha\, m_X^3\frac{m_{\rm DM}}{\rho_c}\sqrt{\frac{3m_p^2}{\rho_{\rm PBH}^{\rm ev}}} \left(\frac{a_{\rm ev}}{a_0}\right)^3 T_{\rm ev}\MeijerG*{2}{1}{1}{3}{-\frac{9}{2}}{-\frac{1}{2}, \frac{1}{2}, -\frac{11}{2}}{\frac{m_X}{2T_{\rm ev}},\frac{1}{2}}\,,\\
\Omega_{\rm IV} &=&\alpha\, m_X^3\frac{m_{\rm DM}}{\rho_c}\frac{36\sqrt{10}}{\pi \sqrt{g_{\star,\rho}(m_X)}}\frac{g_{\star,s}(T_0)}{g_{\star,s}(m_X)} \frac{T_0^3 m_p}{m_X^4}\MeijerG*{2}{1}{1}{3}{1}{\frac{3}{2}, \frac{5}{2}, 0}{\frac{m_X}{T_0},\frac{1}{2}}\,,
\eea
\end{subequations}
\end{widetext}
where $G^{m,n}_{p,q}$ are the Meijer's G functions and $\rho_c=3H_0^2m_p^2$ is the critical density. In practice, the argument of the Meijer's functions is larger than unity in each of the different regimes and the following limits can be used
\begin{subequations}
\bea
\MeijerG*{2}{1}{1}{3}{1}{\frac{3}{2}, \frac{5}{2}, 0}{x\gg 1,\frac{1}{2}}&=&\frac{3\pi}{8}\,,\\
\MeijerG*{2}{1}{1}{3}{-\frac{3}{4}}{-\frac{1}{2}, \frac{1}{2}, -\frac{7}{4}}{x\gg 1,\frac{1}{2}}&\approx & 1.026 \times x^{-7/2} \,,\\
\MeijerG*{2}{1}{1}{3}{-\frac{9}{2}}{-\frac{1}{2}, \frac{1}{2}, -\frac{11}{2}}{x\gg1,\frac{1}{2}}&\approx & 2880 \times x^{-11}\,.
\eea
\end{subequations}

Finally, it is important to note that in Eq.~\eqref{eq:ev}, the choice of the parameter $\nu$ can lead to significant variations of the evaporation temperature, which can have a substantial impact on the estimation of the relic density. Moreover, in reality, the rate $\Gamma_{\rm PBH}$ is not constant and increases during the evaporation process. This typically leads to a slight increase in the SM temperature towards the end of the evaporation process. Therefore, the energy density of the SM bath in Regime III can be overestimated if the scaling relation $T\propto a^{-3/8}$ is assumed using the correct value of the evaporation temperature. To take this deviation into account, we first fix $\nu=0.45$ in order to compute the effective value of $a_{\rm ev}$ and $T_{\rm ev}$ and check that this choice reproduces well the numerical results for our choice of parameters. Then, in Regime III, while we keep fixed the value of $a_{\rm ev}$, we re-adjust our choice to $\nu=0.9$ in order to calculate the value of $T_{\rm ev}$ used in the expression of $\Omega_{\rm III} h^2$. In this way, the temperature of the SM bath during the FI production is comparatively lower than it would be if the scaling relation $T\propto a^{-3/8}$ would remain true up to the real evaporation temperature.

\subsection{Freeze-Out Mechanism}

The case of Freeze-Out was treated in the context where an arbitrary sector of equation-of-state parameter $w$ reheats the universe~\cite{Arias:2019uol}. In our case, PBHs behave mainly like matter during most of the Universe's history, corresponding to the case $w=0$. However, as mentioned above, special care needs to be adopted when discussing Regime III as the evolution of $T(a)$ might not behave like a power law at the end of the evaporation process. 

Following Ref.~\cite{Arias:2019uol}, the FO production can be described by using the non-relativistic cross-section $\langle\sigma v\rangle$ defined in Eq.~\eqref{eq:sigmav} and defining $N_{\rm DM}=a^3 n_{\rm DM}$
\be
\frac{\dd N_{\rm DM}}{\dd a} = -\frac{\langle\sigma v\rangle}{H a^4}\left(N_{\rm DM}^2-N_{\rm DM,\, eq}^2\right)\,.
\ee
Defining $x_{\rm FO}\equiv m_{\rm DM}/T_{\rm FO}$ and 
\be
\kappa = \frac{30}{\pi^2}\frac{\rho_{\rm PBH}^{\rm in}}{m_{\rm DM}}\frac{1}{g_\star(T_{\rm in})T_{\rm in}^3}\,,
\ee
one obtains
\begin{itemize}
  \item Regime I and IV: \be
x_{\rm FO}=\ln\left[\frac{3}{2}\sqrt{\frac{5}{\pi^5 g_\star(T_{\rm FO})}}g_{\rm DM}m_{\rm DM}m_p\langle\sigma v\rangle\sqrt{x_{\rm FO}}\right]\,,
\ee
\item Regime II:
\be
x_{\rm FO}=\ln\left[\frac{3}{2}\sqrt{\frac{5}{\pi^5 g_\star(T_{\rm FO})}}\frac{g_{\rm DM}m_{\rm DM}m_p\langle\sigma v\rangle}{\sqrt{\kappa}}\right]\,,
\ee
\item Regime III:
\be
x_{\rm FO}=\ln\left[\frac{3}{2}\sqrt{\frac{5}{\pi^5 g_\star(T_{\rm FO})}}\frac{g_{\rm DM}m_p\langle\sigma v\rangle}{m_{\rm DM}}T_{\rm ev}^2 x_{\rm FO}^{5/2}\right]\,.
\ee
\end{itemize}
With those definitions of $x_{\rm FO}$, one can infer the value of the relic density in each of the regimes as
\bea
\Omega_{\rm I} &=&\frac{15}{2\pi}\frac{x_{\rm FO}}{\sqrt{10 g_\star(T_{\rm FO})}}\frac{s_{\rm eq}}{m_p\langle\sigma v\rangle\rho_c}\left(\frac{a_{\rm eq}}{a_0}\right)^3\,,\nonumber\\
\Omega_{\rm II}&=&\frac{45}{4\pi}\frac{1}{m_{\rm DM}m_p\langle\sigma v\rangle}\sqrt{\frac{\kappa}{10g_\star(T_{\rm FO}}}x_{\rm FO}^{3/2}\,,\nonumber\\
\Omega_{\rm III}&=&\frac{\pi}{2}\sqrt{\frac{g_\star(T_{\rm FO})}{10}}\frac{m_{\rm DM}^2}{m_p\langle\sigma v\rangle}\kappa\left(\frac{m_{\rm DM}T_{\rm ev}}{T_{\rm FO}^2}\right)^2\,,\nonumber\\
\Omega_{\rm IV} &=&\frac{15}{2\pi}\frac{x_{\rm FO}}{\sqrt{10 g_\star(T_{\rm FO})}}\frac{s_0}{m_p\langle\sigma v\rangle\rho_c}\,,\nonumber\\
\eea
where $s_0=\frac{2\pi^2}{45}g_\star(T_0)T_0^3$ and $s_{\rm eq}=\frac{2\pi^2}{45}g_\star(T_{\rm eq})T_{\rm eq}^3$.

\section{Constraints}\label{sec:constraints}
In this Section, we review the different constraints that we consider when solving the Boltzmann equations numerically.

\subsection{Thermalization of Evaporation Products}

As described in details in Ref.~\cite{paper1}, the DM and mediator particles produced via PBH evaporation follows a non-trivial phase-space density distribution that is dictated by the dynamics of the evaporation. In particular, for PBHs with a large Hawking temperature, those evaporation products can be significantly boosted when emitted. For an arbitrary species, $i$, let us introduce the phase-space distribution of the particle of that species right after evaporation as
\be
f_{\rm ev}(p, t_{\rm ev})\equiv \frac{\dd \mathcal N_{s_i}}{\dd p}(p)\,.
\ee
After evaporation, unless it is affected by subsequent collision processes, this distribution is simply redshifted as follows:
\be
f_{\rm ev}(p, t_{\rm ev})\ \longrightarrow\ f_{\rm ev}(p, t)\equiv f_{\rm ev}\left(\frac{a(t)}{a(t_{\rm ev})}\times p,t_{\rm ev}\right)\,.
\ee
When they are emitted, DM or mediator particles may not interact immediately with the SM plasma. However, because their phase-space density distribution evolves as the Universe expands, the corresponding interaction rates might vary and eventually lead these particles to interact efficiently with the thermal bath or with the preexisting relic abundance of DM particles produced via FO or FI.

\begin{figure}[t!]
 \centering
 \includegraphics[width=\linewidth]{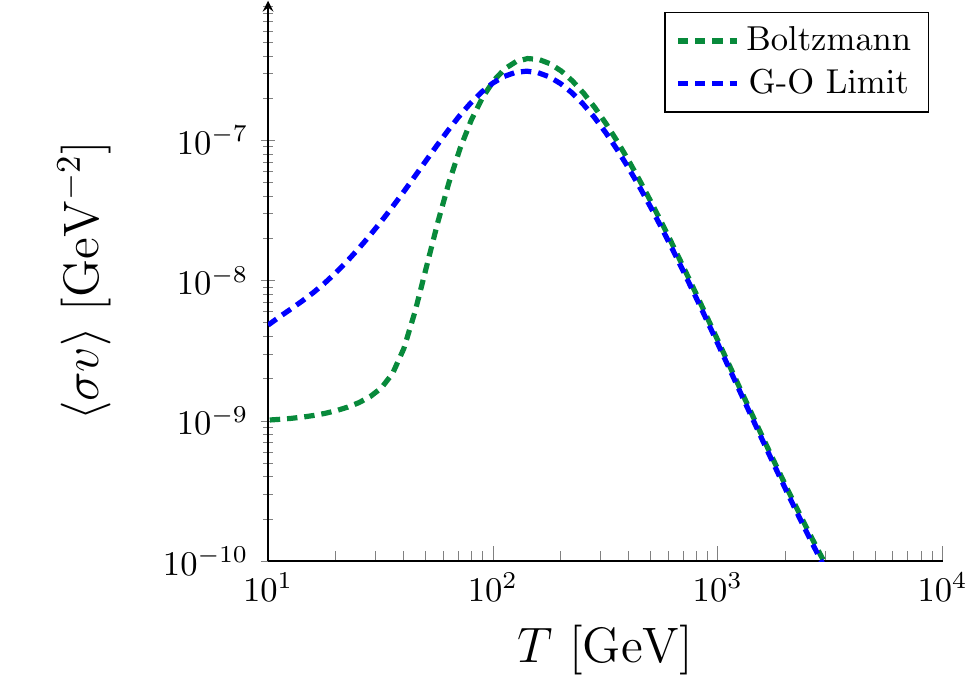}
 \caption{\label{fig:test_sigma}\footnotesize Comparison of cross-section for PBH produced DM to annihilate into SM particles as a function of temperature. The blue (green) dashed line shows the cross-section   computed using the phase-space distributions of PBH-emitted  particles  in the geometric optic limit (Boltzmann distribution) for $\mDM=10^2\mathrm{GeV}$, $m_X=10^3\mathrm{GeV}$, $g_V=g_D=0.73$.}
\end{figure}

Let us consider a particle species ``1'' (DM or mediator) produced from evaporation and study its scattering with particle species ``2'' which is thermalized with the SM bath. At a given time, we denote by $f_{\rm ev}(p_1)$ and $f_{\rm th}(p_2)$ their respective phase-space density distributions in momentum space. We also denote the amplitude of the scattering 1+2 $\to$ 3+4 by $\mathcal M_{1+2\to 3+4}$, regardless of what the final states are. Immediately after PBH evaporation, the interaction rate between the two populations of particles leading to a depletion or momentum transfer of the evaporated species can be evaluated as
\begin{widetext}
\be
g_1 \int C[f_{\rm ev}]\frac{\dd^3 p_1}{(2\pi)^3} \sim -\int f_{\rm ev}(p_1)f_{\rm th}(p_2)\overline{\left|\mathcal M_{1+2\to 3+4}\right|^2}\times \delta^4(p_1+p_2-p_3-p_4)\dd\Pi_1\dd\Pi_2\dd\Pi_3\dd\Pi_4\,.
\ee
\end{widetext}

In order to estimate the efficiency of such a scattering process after evaporation, it is convenient to define an annihilation cross-section that is averaged over both the thermal and {\em evaporated} distributions as follows
\be \label{eq:thevav}
\langle\sigma \cdot v\rangle_{\rm th+ev}\equiv\frac{\displaystyle\int \sigma \cdot v f_{\rm ev} f_{\rm th} d^{3} \vec{p}_1 d^{3} \vec{p}_2}{\left[\displaystyle\int d^{3} \vec{p}_1 f_{\rm ev}\right]\left[\displaystyle\int d^{3} \vec{p}_2 f_{\rm th}\right]}\,.
\ee
 Given the definition in Eq.~\ref{eq:thevav}, one can estimate the scattering efficiency of an evaporated particle scattering on a thermal particle by evaluating the interaction rate
\be
\Gamma_{\rm ev\to th}\equiv \frac{\langle\sigma \cdot v\rangle_{\rm th+ev} \times n^{\rm th}}{H}\,.
\ee
Similarly, one can estimate the ability of evaporated particles to self-scatter by evaluating rates of the form
\be
\Gamma_{\rm ev\to ev}\equiv \frac{\langle\sigma \cdot v\rangle_{\rm ev+ev} \times n^{\rm ev}}{H}\,.
\ee
In our calculation, in order to ensure that thermalization does not take place --- both in the FI mechanism and in the FO mechanism when evaporation takes place after the FO ---  we follow the following procedure: First, we compute the contribution to the DM energy density produced through evaporation and compare it to the value of the DM energy density produced via FO at the time of PBH evaporation. If this contribution does not constitute more than 10\% of the total DM energy budget at the time of evaporation, we consider that later interactions will not affect the value of the final relic abundance by more than a few percents and our derivation remains valid in this case. Suppose this contribution exceeds 10\% at the time of evaporation. In that case, we demand that the interaction rate of DM and mediator particles with any other particles is smaller than unity to ensure that no thermalization of the evaporation products happens, all the way from the time of evaporation to present time.

In practice, the full numerical integration of the different amplitudes over the evaporation phase-space distribution is time and resource-consuming. Therefore, it is convenient to approximate this distribution by a Boltzmann distribution to perform at least part of the integration analytically. This treatment is described in detail in Appendix \ref{app:XsectionsdiffT} where we provide the annihilation cross-section averaged over two Boltzmann distributions with different temperatures. However, as shown in Ref.~\cite{paper1}, approximating the distribution of evaporated particles by a Boltzmann distribution can be erroneous, especially for particles of momentum $p\gg \TBH,\mDM$. In Fig.~\ref{fig:test_sigma} we illustrate such an error by comparing the value of the cross-section of DM annihilation into SM model fermions $\langle\sigma v\rangle_{\rm ev+ev}$ computed with the evaporation distribution in the geometrical-optics limit (blue dashed line) \cite{paper1, Baldes:2020nuv} and computed with a Boltzmann approximated distribution (green dashed line). 
{As one can see from this figure, for $T\gg m_X$, the Boltzmann approximation turns out to be a good approximation. However, at a temperature $T\lesssim m_X$ the $s$-channel resonance,  the Boltzmann approximation leads to a slight underestimation of the interaction rate.} This can be understood by inspecting the shape of the evaporation distribution as compared to the Boltzmann distribution. Indeed, as it was shown in Refs.~\cite{paper1, Baldes:2020nuv}, although the Boltzmann approximation is a good approximation around the peak of the distribution, the tail of the evaporation distribution at large momenta is significantly larger than in the Boltzmann approximation. Therefore, at a temperature $T \lesssim m_X$, the resonance of the distribution stands at larger momenta $p\gtrsim T$ for which the evaporation distribution is larger, explaining the excess. For the sake of evaluating whether thermalization occurs, using the Boltzmann approximation is sufficient since the peak of the thermally-averaged cross-section is unaffected. For a $t$-channel annihilation, the calculation is not affected by the presence of any resonance, and the Boltzmann approximation turns out to be very acceptable.

\subsection{Warm/Non-Cold Dark Matter}
The dark matter particles which are produced through evaporation carry a significant momentum at the time of evaporation. After that, although they lose part of their kinetic energy via gravitational redshift, DM particles may still be quasi-relativistic at the time of structure formation and erase structures on scales below its free-streaming length. The study of the Lyman-$\alpha$ forest provides one of the strongest constraints on warm or non-cold DM candidates (see e.g. \cite{Hui:1996fh,Gnedin:2001wg}). In Ref.~\cite{Boyarsky:2008xj, Baur:2017stq} such a constraint was derived assuming that only a fraction of the DM relic abundance is warm. In order to exploit such a constraint in our scenario, we calculated the fraction of the DM relic abundance produced via the evaporation of PBHs and calculated their average momentum $\langle p\rangle_{\rm ev}$ at the time of evaporation using the results derived in Ref.~\cite{paper1}. The velocity of the DM particles today is obtained by simply redshifting the value of their average momentum at the time of evaporation as
\be
v_{0} = \frac{a_{\rm ev}}{a_0}\times\frac{\langle p\rangle_{\rm ev}}{m_{\rm DM}}\,.
\ee
Defining the fraction of evaporated particles as
\be
\eta_{\rm ev}=\frac{\Omega_{\rm DM,\, ev}}{\Omega_{\rm DM,\, tot}}\,,
\ee
we can then use the constraints derived in Ref.~\cite{Boyarsky:2008xj,Baur:2017stq} to determine whether a given point of the parameter space is excluded from Lyman-$\alpha$ measurements\footnote{Note that in Ref.~\cite{Baur:2017stq} such a constraint is provided on the mass of a thermal warm DM particle $m_{\rm WDM}$. Following Ref.~\cite{Baldes:2020nuv}, we simply use that $v_{\rm WDM}\approx 3.9\times 10^{-8}(\mathrm{keV}/m_{\rm WDM})^{4/3}$ in order to extract a constraint on the velocity $v_0$.}. In particular, we note that such constraints allow the possibility that a fraction $\eta_{\rm ev}\lesssim 0.02$ is warm today. Thus, regions of the parameter space in which PBHs produce less than 2\% of the relic density are not affected by the warm dark matter constraint derived from Lyman-$\alpha$ measurements. {Note that a more refined analysis of the WDM constraint has been achieved in Refs.~\cite{Baldes:2020nuv, Auffinger:2020afu} by studying the matter distribution “transfer
function” using the \textsf{CLASS} code. The results sketched in those studies in the context of mixed warm-cold DM scenarios agree qualitatively with our findings. We leave the adaptation of this technique to our scenario for future work.}

\section{Results}\label{sec:results}
\begin{figure}[t!]
\centering
 \includegraphics[width=\linewidth]{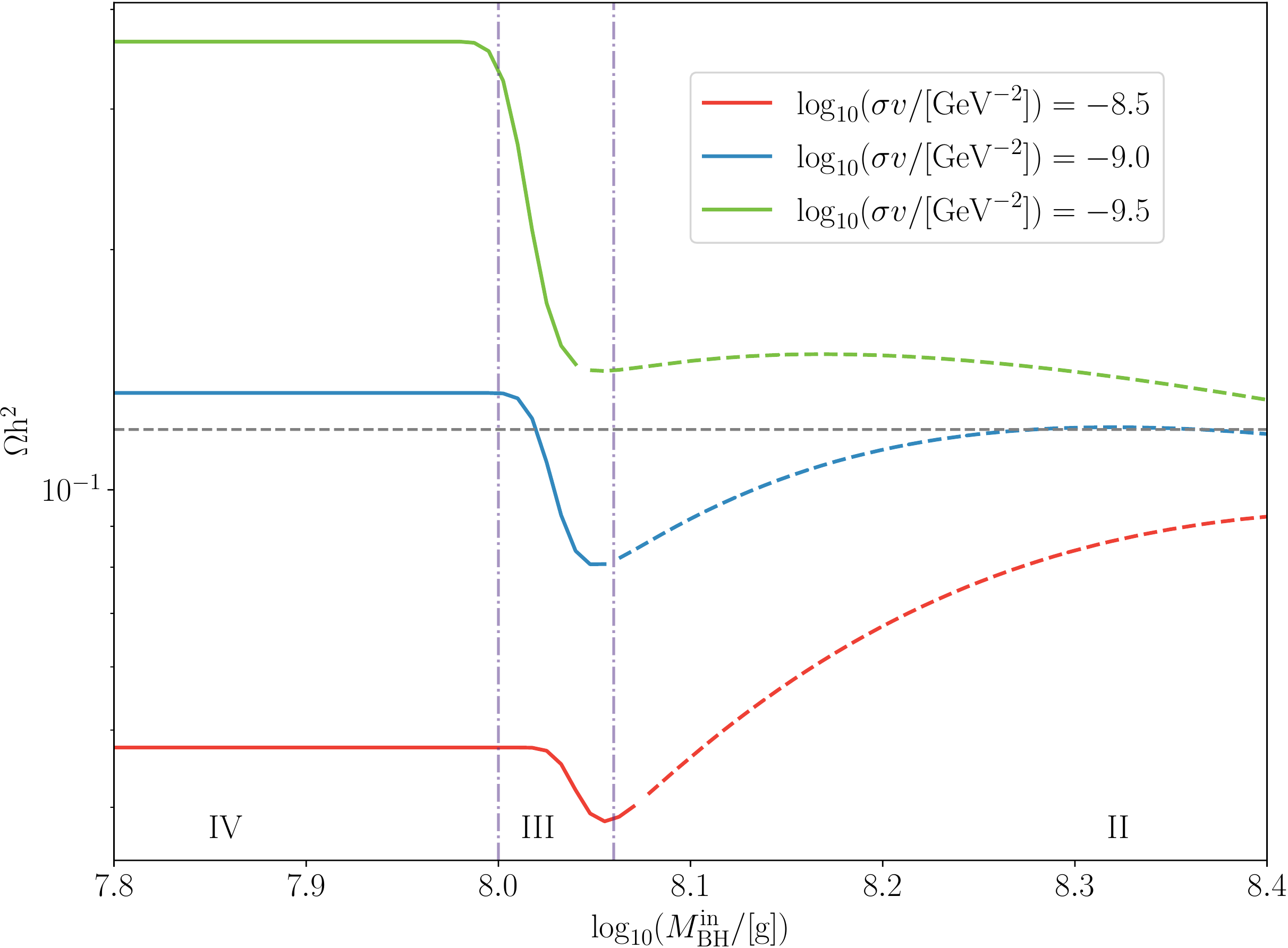}
 \caption{\label{fig:FO-therm}The relic density as a function of the initial PBH mass for the FO mechanism with $\beta^\prime = 10^{-10}$,  $m_X = 10\,\rm{GeV}$, $\mDM = 1\, \rm{GeV}$,  $\mathrm{Br}(X\to \mathrm{DM}) = 0.5$ and $a^{*}=0$ for
thermally averaged cross-section values $\log_{10}(\sigma v/[\text{GeV}^{-2}]) =  -8.5, -9.0, -9.5$ shown in red, blue and green  respectively. The grey dashed line indicates the observed relic density value. The grey dot-dashed lines indicate the separation of regimes IV, III and II.}
\end{figure}
\begin{figure*}
    \centering
    \includegraphics[width=0.7\linewidth]{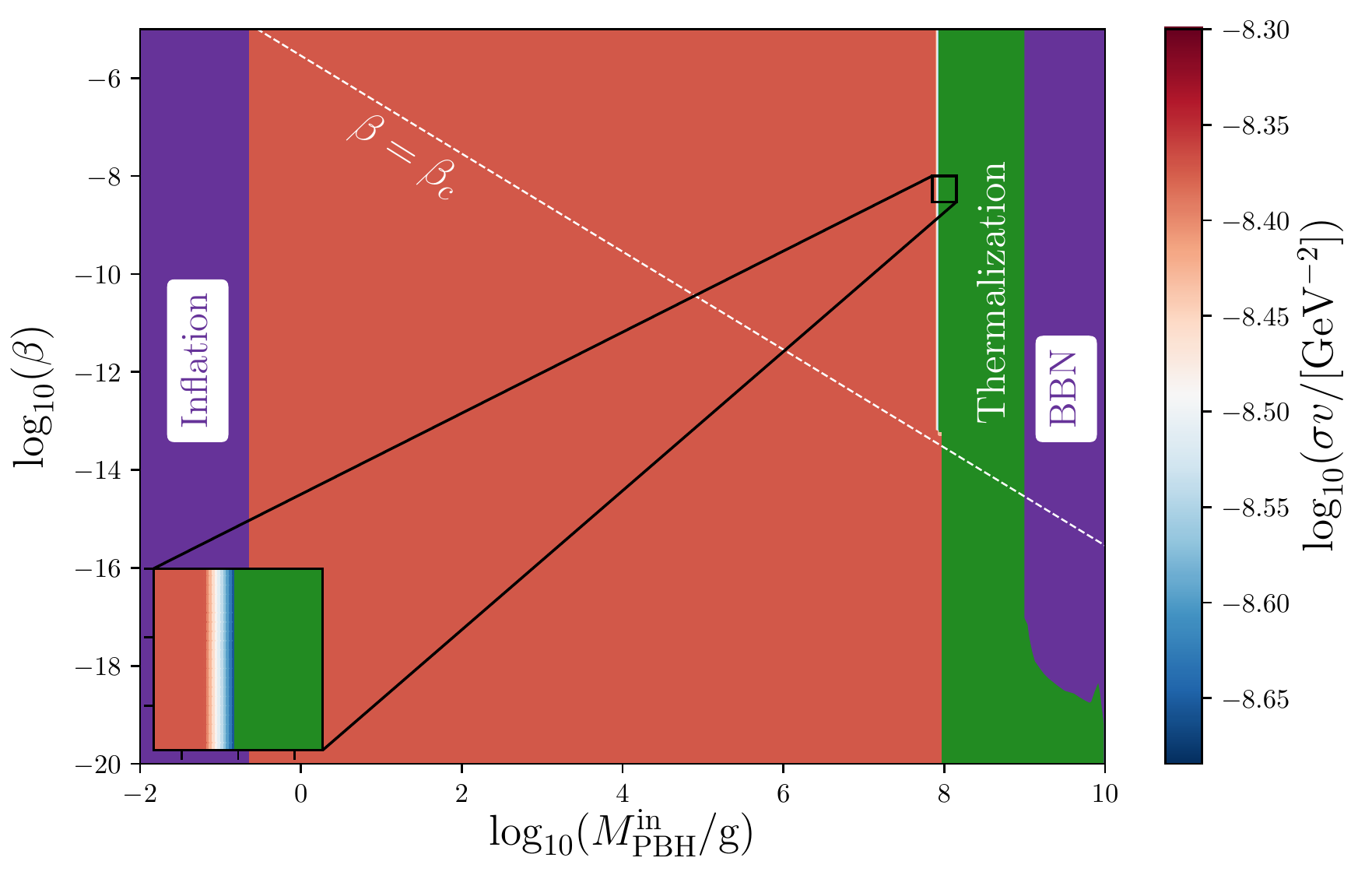}
    \caption{\label{fig:scan_FO}\footnotesize Two-dimensional scan over the PBH fraction $\beta$ and mass $\MBH$ for a mediator mass $m_X=10\,\mathrm{GeV}$ and a dark matter mass $m_{\rm DM}=1\,\mathrm{GeV}$, and $\mathrm{Br}(X\to \mathrm{DM})=0.5$. The color map indicates the value of the non-relativistic cross-section of DM annihilation leading to the correct relic abundance in the Freeze-Out case. See the main text for a description of the different constraints.}
\end{figure*}
\subsection{Freeze-Out Regime}\label{sec:FO}
As discussed in \secref{sec:thermevap},  thermalization  between the thermally produced DM 
 and the evaporation products of the PBHs  tends to occur due to the large annihilation cross-sections
typical of the FO mechanism.
Nonetheless, in this short Section, we study the effect PBH evaporation can have on FO and identify regions
of the parameter space where dilution of the DM relic abundance can occur due to PBH evaporation. We also highlight a regime of thermalization where a more careful treatment involving evolving the phase-space distributions of the DM should be used.

As we have seen in Sec.~\ref{sec:interplay}, the evaporation of PBHs may play an important role in modifying the FO of DM particles for a DM mass $\mDM\lesssim 1 \mathrm{GeV}$. In that regime of parameter space, an annihilation cross-section of order $10^{-9}\mathrm{GeV}^{-2}$ together with couplings $g_V,g_D \sim \mathcal O(0.1)$ require a mediator of mass $m_X\sim \mathcal O(10)\mathrm{GeV}$. Such a possibility clearly has problems avoiding other experimental constraints. Any boson of this mass interacting with SM quarks or leptons would be present in collider and fixed target experiments. A potential way around is if the mediator couples only to light (sterile) neutrinos or to some other neutral fermions that remain in equilibrium with the SM when $T\lesssim \mDM$. Furthermore, indirect detection constraints could be avoided if $\langle \sigma v\rangle$ is velocity suppressed, this can be achieve if our DM is Majorana. Since this work is primarily focused on the interplay between thermal DM processes and PBHs, we do not consider these constraints further.

\begin{figure*}[t!]
\centering
 \includegraphics[width=\textwidth]{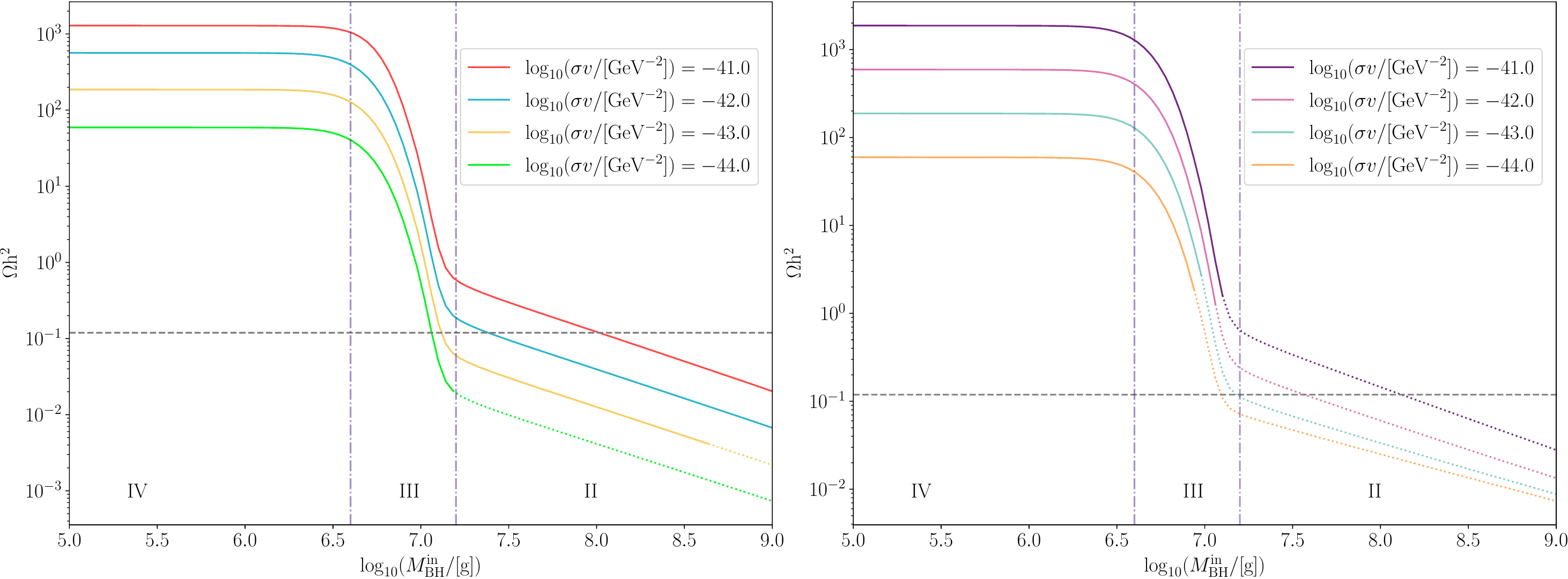}
 \caption{\label{fig:FI_dilute1} The left (right) plot shows the value of the relic density as a function of the initial 
 PBH mass for the FI mechanism with $\beta^\prime = 10^{-10}$, 
$m_X = 10\,\rm{GeV}$, $\mDM = 10^{-3.0}\, \rm{GeV}$ ($\mDM = 10^{-1.0}\, \rm{GeV}$) and $\mathrm{Br}(X\to \mathrm{DM}) = 0.5$, $a^{*}=0$ for
thermally averaged cross-section values of $10^{-41}$, $10^{-42}$, $10^{-43}$ and $10^{-44}$ $\text{GeV}^{-2}$ shown in red, blue, yellow and green (purple, pink, cyan and orange) respectively. {The dotted portions of the different lines indicate regions where the DM is hot and would disrupt structure formation.} The grey dashed line indicates the observed relic density value. The grey dot-dashed lines indicate the separation of regimes IV, III and II.}
\end{figure*}

In \figref{fig:FO-therm}, we show how the relic abundance changes as a function of the initial PBH mass for  the point $\beta^\prime = 10^{-10}$, $m_X = 10\,\rm{GeV}$, $\mathrm{Br}(X\to \mathrm{DM}) = 0.5$ and  $\mDM = 1.0\, \rm{GeV}$  for $\log_{10}(\sigma v/[\text{GeV}^{-2}]) = -8.5, -9.0, -9.5$  as indicated in red,  blue and green respectively. The region where post-FO DM scattering processes remain active after evaporation and our calculation is no longer valid are shown by dashed lines. From \figref{fig:FO-therm},  we observe that for  $\log_{10}(\MBH^{\rm in}/[\rm{g}]) \lesssim 8.0$, the PBHs have evaporated before FO occurs (corresponding to regime IV), around $T\sim \mDM$, and the relic abundance is unchanged by the presence of the PBHs.  For $\log_{10}(\MBH^{\rm in}/[\rm{g}]) \sim 8.0$, evaporation occurs \emph{during} FO (corresponding to regime III) and the entropy injection from the PBH evaporation dilutes  the relic density by approximately $80\%$ independent of the annihilation cross-section. For heavier initial PBH masses, the Universe undergoes a stage of matter domination corresponding to regime II, but in this regime DM particles emitted by PBHs can actively scatter with other SM and/or DM particles, which invalidates our no-thermalization assumption. Naturally, the initial mass of the PBH, which triggers thermalization will change depending on the mass of the DM freezing out: the heavier the DM,  the earlier the FO occurs and therefore, the lighter the PBHs would have to be to provide the dilutionary effect and thermalization.
We found that for larger annihilation cross-sections, $\log_{10}(\sigma v/[\text{GeV}^{-2}]) \gtrsim -8.0$,  thermalization always occurs for this given point.   

In Fig.~\ref{fig:scan_FO} we consider the same benchmark point as in Fig.~\ref{fig:FO-therm} and scan over the PBH mass, energy fraction $\beta$, and annihilation cross-section, using our analytical expressions in order to estimate the FO contribution to the relic abundance. The color map stands for the value of the cross-section leading to the correct abundance. The green-shaded area labelled {\em Thermalization} stands for the region where the evaporation takes place after FO and where the DM particles produced 
by PBH evaporation are expected to efficiently interact with the other DM/SM particles. As we described in Sec.~\ref{sec:constraints}  solving the Boltzmann equation for the full DM phase-space distribution is required in this region of parameter space which goes beyond our present treatment. As one can see, there exists a narrow region of the parameter space where PBHs contribute to a small fraction of the relic density, therefore requiring a smaller annihilation cross-section in order to match with observations. At smaller values of $\MBH$ the evaporation takes place before FO and thermalization processes wash out the contribution of PBHs to the relic density. We insist on the fact that this structure of the parameter space is generic to DM masses lighter than $\lesssim \mathcal{O}(\mathrm{GeV})$. For that reason, there exists practically no regime in which an early PBH domination era can significantly affect the FO dynamics while escaping the post-FO thermalization constraint. Indeed, when the PBH domination is able to deplete the contribution of the FO to the relic density, the evaporation products are always sufficiently boosted in order to interact with the SM bath after evaporation. Finally, regimes with larger DM masses in which PBHs do not dominate the energy density of the universe could in principle lead to a sizeable contribution of PBH evaporation to relic abundance. However, in that case, this contribution can only be negligible in order to avoid the warm DM constraint as was shown in Ref.~\cite{Masina:2020xhk, Masina:2021zpu}.

\subsection{Freeze-In Regime}\label{sec:FI}

Due to the small cross-sections typical of the FI mechanism, thermalization between the population of thermally produced and PBH produced DM is not likely to occur and we expect that dilution may play an important role in certain regions of the parameter space. However, to ensure that thermalization does
not occur, we check that the thermal cross-section (computed using the narrow-width approximation) multiplied by the number density of thermal DM particles does not exceed the Hubble scale at any point in the temperature evolution.

We have identified a region of the parameter space where the FI mechanism can be affected by the late time evaporation of the PBH population. In particular, this region lies in the left green-shaded triangle of the right plot of \figref{fig:MBHmin}. The effect of PBH evaporation on FI is shown in \figref{fig:FI_dilute1} where the relic density is plotted as a function of the initial PBH mass for the two generic points which exhibits  PBH domination:  $\beta^\prime = 10^{-10}$, 
$m_X = 10\,\rm{GeV}$, $\mathrm{Br}(X\to \mathrm{DM}) = 0.5$ and  $\mDM = 10^{-3.0}\, \rm{GeV}$ ($\mDM = 0.1\, \rm{GeV}$)  for the left (right) plot. The solid coloured lines indicate when all relevant tests are passed, i.e. the DM is not hot and does not thermalize at or after evaporation. While the dotted coloured lines indicate regions where the DM is hot and would disrupt structure formation  \cite{Boyarsky:2008xj}.

As we would expect, in both cases, the relic density decreases as the cross-section is reduced from $10^{-41}$  to $10^{-44}\,\rm{GeV}^{-2}$. Further, for $\log_{10}(\MBH^{\rm in}/[\rm{g}]) \gtrsim 7$, the late time entropy injection (regime III) from large mass PBHs  dilutes the relic density. This dilutionary effect is more significant for smaller cross-sections as there is less DM to dilute to begin with. Interestingly, the late-time dilution offers the possibility for certain points in the parameter space, which would otherwise overproduce DM, to be consistent with the observed DM abundance.  For smaller cross-sections (shown in yellow and green of the left plot of  \figref{fig:FI_dilute1}) and larger masses of PBHs, the relative contribution of DM produced by the PBHs compared with DM produced from the Freeze-In is larger, and therefore the hot DM constraints exclude these scenarios.
We find that for larger DM masses i.e.  $\mDM = 10^{-1.0}\, \rm{GeV}$, as shown in the right plot of \figref{fig:FI_dilute1}, larger initial PBH $\log_{10}(\MBH^{\rm in}/[\rm{g}]) \gtrsim 7$ provide the same dilution but tend to produce DM which is too hot  regardless of the annihilation cross-section. The logic follows from \equaref{eq:sigmav} where we observe that for a fixed annihilation cross-section, branching ratio and mediator mass, the values of the visible ($g_V$) and dark sector couplings ($g_D$) increase as a function of the DM mass.  Therefore, the 
larger the dark matter mass, the larger the proportion of DM production by the PBHs (since $\Gamma_{X\to \mathrm{DM}}$ increases quadratically in $g_D$, see \equaref{eq:GamXDM}) and therefore, the warm dark matter constraint is more easily violated for heavier DM, for this set of parameters.

\begin{figure}[t!]
\centering
 \includegraphics[width=\linewidth]{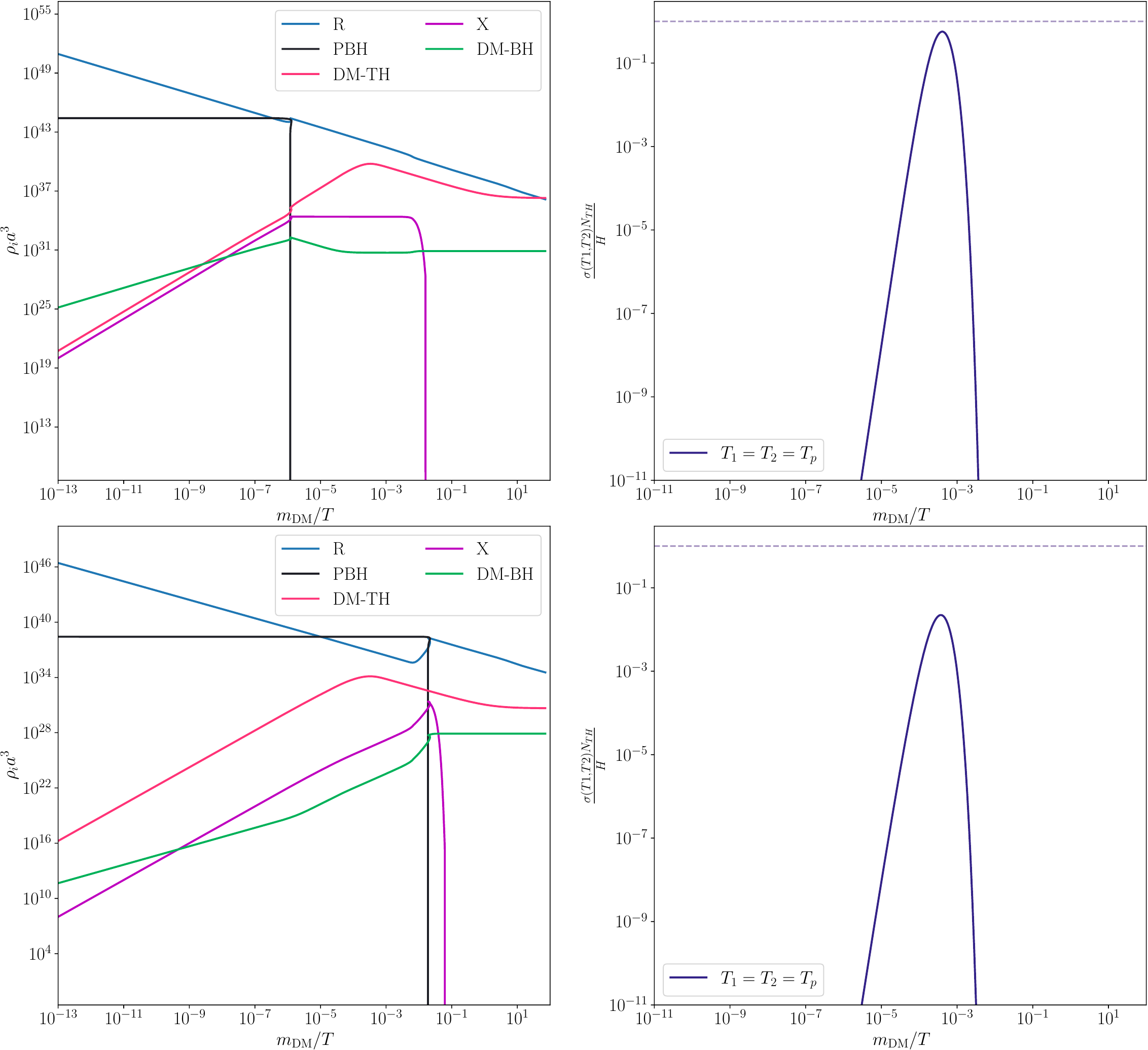}
 \caption{\label{fig:FI_dilute2} Top (bottom) left shows the time  
 evolution of the energy density of the radiation, PBHs, thermally produced DM, PBH produced DM and mediator is shown in blue, black, green, red and purple respectively for $M_{\rm{PBH}}=10^{5}$ g ($M_{\rm{PBH}}=10^{8}$ g). The top (bottom) right shows the thermal cross-section multiplied by the number density of thermal DM particles  divided by the Hubble expansion rate as a function of inverse temperature. The grey dashed line indicates the $\sigma N_{TH}/H =1 $.}
\end{figure}

As smaller DM masses are less susceptible to the hot DM constraint, for these choices of parameters,  we focus on the case $\mDM = 10^{-3.0}\, \rm{GeV}$
and study it in further depth.  In particular, for $\log_{10}(\sigma v/[\text{GeV}^{-2}]) =-41.0$, we plot the evolution of the radiation (blue), PBH (black), thermally (red), PBH produced DM (green) and mediator (purple)
energy density for $\log_{10}(\MBH^{\rm in}/[\rm{g}]) = 5.0$ ($\log_{10}(\MBH^{\rm in}/[\rm{g}]) = 8.0$) in the top left (bottom) plot of \figref{fig:FI_dilute2}.
 The plots on the right of \figref{fig:FI_dilute2}, demonstrate that these points of the parameter space never experience thermalization. Hence the two populations of DM (thermally and PBH produced) do not interact and can be treated separately.
For $\log_{10}(\MBH^{\rm in}/[\rm{g}]) = 5.0$ (top plot), we observe that there is a short period of PBH domination and  evaporation  \emph{before} freeze at $T \sim M_X$. Prior to evaporation, the PBH-produced DM (shown in green) is increasing in abundance.
 We note that the contribution of the thermally produced DM far exceeds the PBH contribution. 
We can contrast this scenario with the $\log_{10}(\MBH^{\rm in}/[\rm{g}]) = 8.0$ case (bottom plot) where we observe that 
PBH evaporation occurs \emph{after} Freeze-In. As the initial PBH mass is larger and the value of $\beta$ is fixed, there is a greater 
PBH domination than in the previous case. The effect of the entropy dump
(as seen in the change in gradient of the radiation energy density shown in blue) is significant enough to dilute the relic density. Comparing the top and bottom right plots of \figref{fig:FI_dilute2}, we find that the temperature evolution of $\sigma v N_{TH}/H$
is also affected. While the cross-section is unaffected, the number of thermally produced dark matter in the plasma is less
in the case of late-time evaporation (comparing the pink lines of the top and bottom plot of \figref{fig:FI_dilute2}), and due to the reheating of the Universe, the Hubble rate is larger. Hence the suppression of $\sigma v N_{TH}/H$ in the late-time evaporation case. 

\begin{figure}[t!]
\centering
 \includegraphics[width=\linewidth]{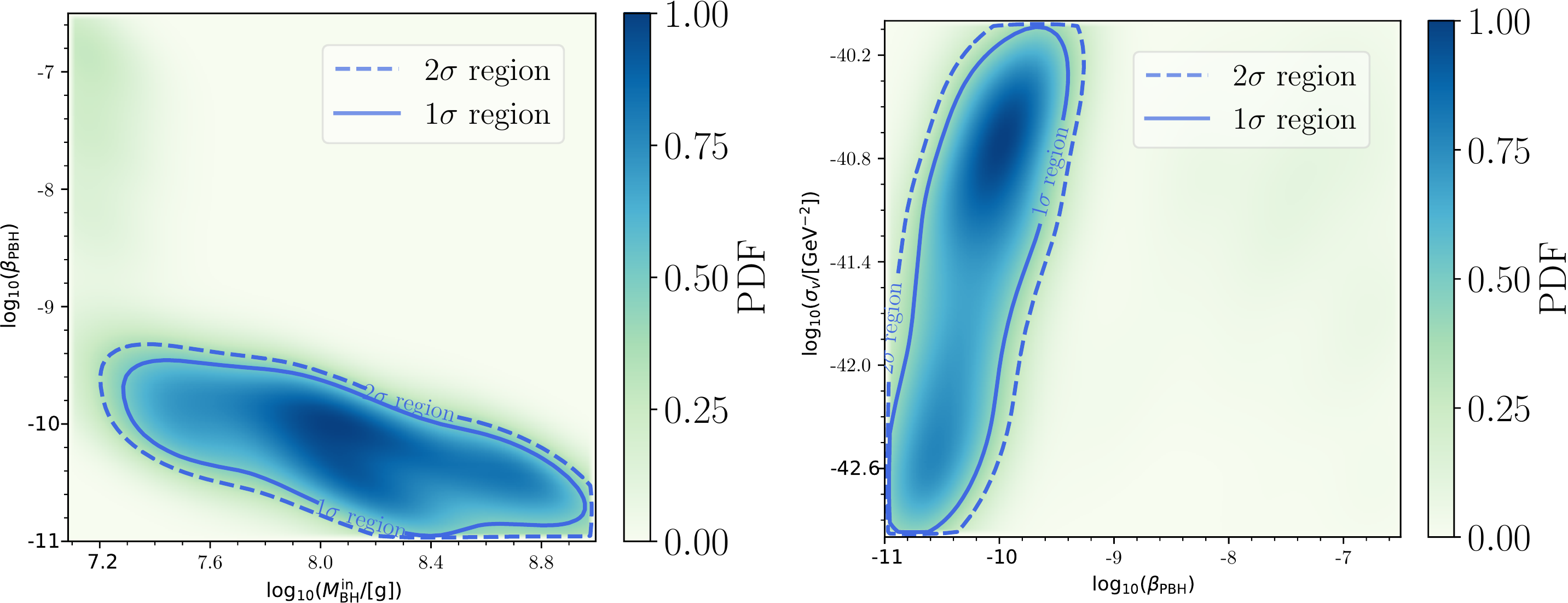}
 \caption{\label{fig:FI_dilute3} Two-dimensional plots showing the 
 correlations between $\log_{10}(\MBH^{\rm in}/[\rm{GeV}])$, $\log_{10}(\beta^\prime)$ and $\log_{10}(\sigma v/[\rm{GeV}^{-2}])$. 
{The darker blue regions shows  higher posterior probabilities and the lighter blue regions show regions with lower posterior probability. The regions of the parameter space, $\vec{p}$, consistent at the one (two) sigma level with the observed relic abundance are within the solid (dashed blue) contours.}
 }
\end{figure}

{For the same point, with fixed $\beta^\prime = 10^{-10}$, 
$m_X = 10\,\rm{GeV}$, $\mDM = 10^{-3.0}\, \rm{GeV}$ and $\mathrm{Br}(X\to \mathrm{DM}) = 0.5$, $a^{*}=0$ we would like to explore the parameter space $\vec{p}$ which we define to be:
$ 5 \leq \log_{10}(\MBH^{\rm in}/[\rm{g}])\leq 9$, $-10 \leq \log_{10}(\beta^\prime) \leq -6.5 $ and $ -43.0\leq \log_{10}(\sigma v/[\rm{GeV}^{-2}]) \leq -40.0$. We would like to determine which regions of $\vec{p}$ are consistent with the observed dark relic abundance at the two sigma level \cite{Aghanim:2018eyx}.
To perform this task, we use {\tt ULYSSES} in conjunction with  {\sc
Multinest}~\cite{Feroz:2008xx,Feroz:2007kg,2013arXiv1306.2144F} (more precisely,
{\sc pyMultiNest}~\cite{pymultinest}, a wrapper around {\sc Multinest} written
in {\sc Python}). The {\sc Multinest} algorithm  provides a
nested sampling algorithm that calculates Bayesian posterior distributions
which we will utilise in order to define regions of confidence.
We place flat priors on the parameter we scan in and {\sc Multinest}  use the log-likelihood  as an objective function:}
\begin{equation}
\log L=-\frac{1}{2}\left(\frac{\Omega \text{h}^2(\vec{p})-\Omega \text{h}^2_{\rm{PDG}}}{\Delta \Omega \text{h}^2}\right)^{2}\,,
\end{equation}
where $\Omega \text{h}^2(\vec{p})$ 
is the calculated relic density for a point in the model parameter space, $\Omega \rm{h}^2_{\rm{PDG}}$ is the best-fit value of the relic density and 
$\Delta \Omega \text{h}^{2}$ is the one-sigma range of the relic abundance \cite{Aghanim:2018eyx}. Once a Multinest run is finished, we use SuperPlot \cite{Fowlie:2016hew} to visualise the posterior projected onto a
two-dimensional plane as shown in \figref{fig:FI_dilute3}. {We note that the darker blue regions shows  higher posterior probabilities and conversely the lighter blue regions show regions with lower posterior probability. The regions of the parameter space, $\vec{p}$, consistent at the one (two) sigma level with the observed relic abundance are within the solid (dashed blue) contours.}

\begin{figure*}
    \centering
    \includegraphics[width=0.7\linewidth]{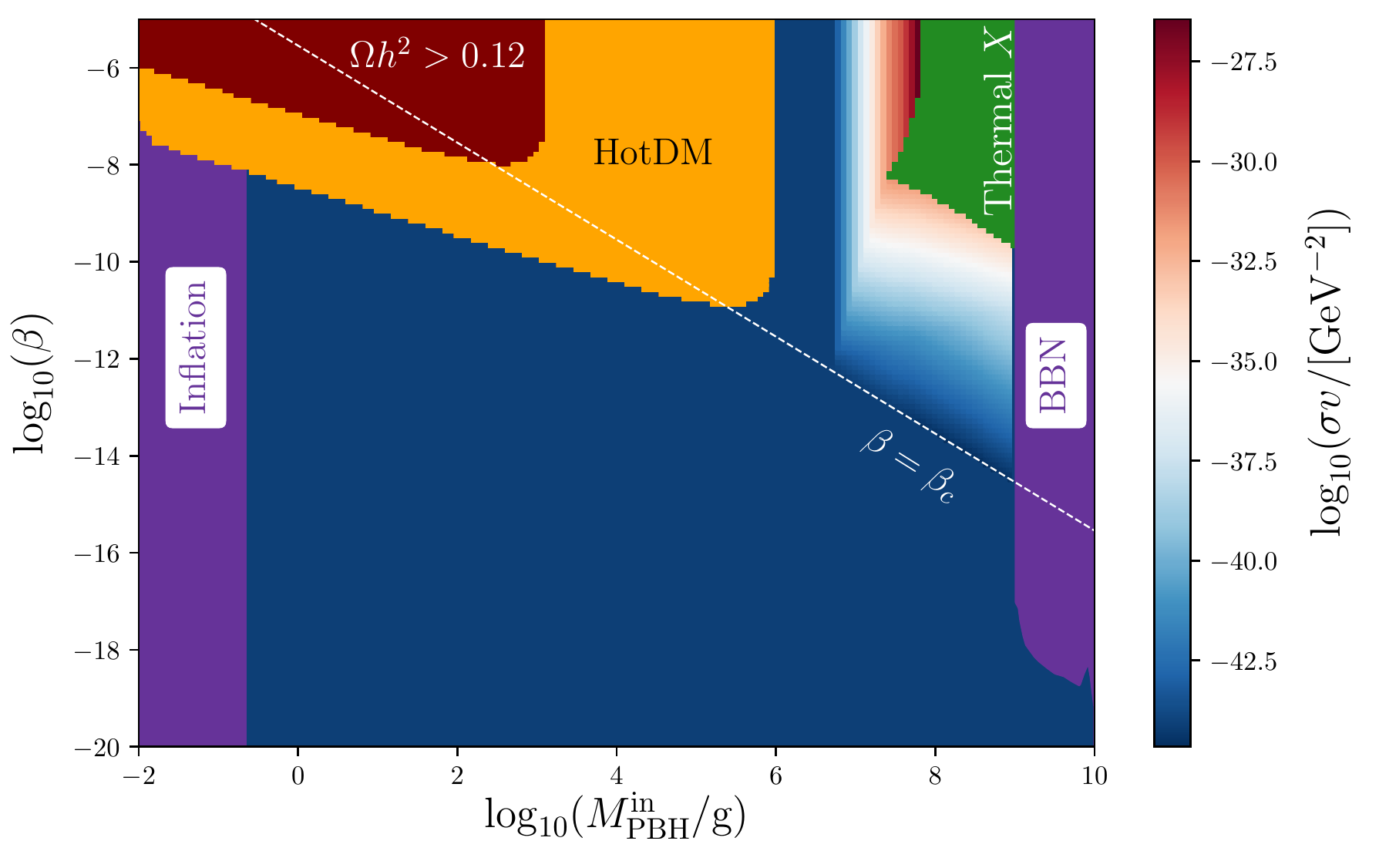}
    \caption{\label{fig:scan_FI}\footnotesize Two-dimensional scan over the PBH fraction $\beta$ and mass $\MBH$ for a mediator mass $m_X=1\,\mathrm{TeV}$, a dark matter mass $m_{\rm DM}=1\,\mathrm{MeV}$, and $\mathrm{Br}(X\to \mathrm{SM})=10^{-7}$. The color map indicates the value of the non-relativistic cross-section of DM annihilation leading to the correct relic abundance in the Freeze-In case. See the main text for a description of the different constraints.}
\end{figure*}
From the left plot of \figref{fig:FI_dilute3}, we find that there is an expected anti-correlation between $\beta^{\prime} $ and $\MBH^{\rm in}$: for larger initial PBH masses, smaller initial fractional energy density of PBHs are required to provide the necessary dilution. Moreover, $\log_{10}(\beta^\prime) \gtrsim -9$ provides too large a dilutionary effect for this point in the parameter space.  From the right plot of \figref{fig:FI_dilute3}, we observe that large cross-sections require larger values of the initial fractional energy density to recover the observed relic density. This is because a larger cross-section produces a larger relic density, requiring a larger entropy dump to dilute it sufficiently.

In \figref{fig:scan_FI} we used our analytical results in order to scan over the DM annihilation cross-section for a given choice of DM and mediator masses to find the value leading to the correct relic abundance. We indicate the region where the evaporation of PBHs leads to an overdensity of DM particles in brown. The orange region is excluded because PBHs produce a significant fraction of warm DM particles, which is excluded by Lyman-$\alpha$ measurements.
The region which transition from dark blue (smaller cross-sections) to red (larger  cross-section) is where the continuous dilution during and after FI from the late time evaporation of the PBHs is effective. Finally, the green area shows where the couplings are large enough to enforce the thermalization of DM and mediator particles which is inconsistent with our Freeze-In scenario. One can see that the region in which PBH dominates the Universe's energy ($\beta>\beta_c$) can lead to a smooth variation of the annihilation cross-section over orders of magnitude as compared to a Freeze-In scenario that would occur in a radiation-dominated background.

\subsection{The effect of varying the mediator mass}
\begin{figure}[t!]
\centering
 \includegraphics[width=\linewidth]{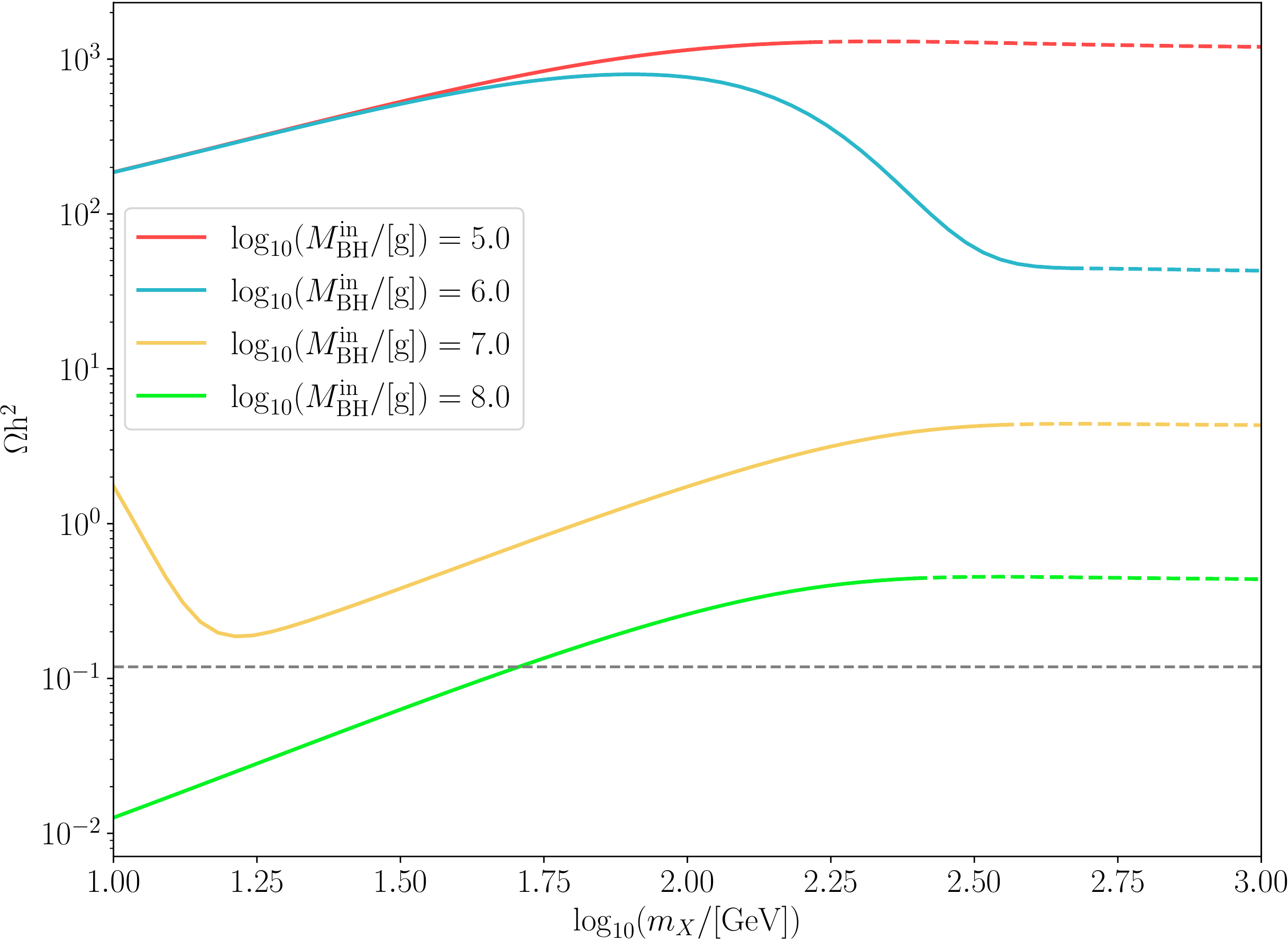}
 \caption{\label{fig:FI_mX}The  relic density as a function of the mediator mass, $m_X$, for the FI mechanism with $\beta^\prime = 10^{-10}$, 
 $\mDM = 10^{-3.0}\, \rm{GeV}$  and $\mathrm{Br}(X\to \mathrm{DM}) = 0.5$, $a^{*}=0$ and $\sigma v = 10^{-41.0}\,\text{GeV}^{-2}$ for
initial PBH mass of  $\log_{10}(\MBH^{\rm in}/[\rm{g}]) = 5.0, 6.0, 7.0, 8.0$ shown in red, cyan, green and purple respectively. The coloured dashed lines indicate when  DM has thermalized and the Freeze-In mechanism transitions to the Freeze-Out mechanism. {The grey dashed line stands for the central value of the DM relic abundance, as measured by the Planck collaboration~\cite{Planck:2018vyg}.}}
 \end{figure}
In \secref{sec:FI}, we studied the effect of varying the initial PBH mass on the FI parameter space where we allowed the DM mass and cross-section to vary. We found that the late-time evaporation of heavy PBHs could sufficiently dilute the relic density such that it is consistent with observation and that for small cross-sections $\log_{10}(\sigma v/ [\text{GeV}^{-2}]) \lesssim -41.0$, lighter DM candidates are favoured, $\mDM \lesssim 10^{-2}\,\text{GeV}$, for that choice of parameters, due to the hot DM constraint.

In the previous section we fixed the mediator mass at $m_X = 1\,\rm{GeV}$. 
In this Section,  we study the effect of varying the mediator mass which is important as this determines the temperature at which the FI occurs. In \figref{fig:FI_mX}, we study the same generic point, $\beta^\prime = 10^{-10}$, 
 $\mDM = 10^{-3.0}\, \rm{GeV}$  and $\mathrm{Br}(X\to \mathrm{DM}) = 0.5$, $a^{*}=0$ and $\log_{10}(\sigma v/ [\text{GeV}^{-2}]) = -41.0$, while varying the mediator mass for fixed initial PBH masses, $\log_{10}(\MBH^{\rm in}/\text{g}) = 5.0, 6.0, 7.0, 8.0$ shown in red, cyan, green and purple respectively. The dashed lines indicate where thermalization takes place and our treatment is no longer valid. 
We see that for smaller PBH masses, the relic abundance is not significantly affected as the mediator mass is varied. This can be seen from the solid red and blue lines of \figref{fig:FI_mX}. This occurs because PBH evaporation occurs \emph{before} Freeze-In for this scenario. However, for all shown values of initial PBH mass, DM thermalization occurs for larger mediator masses, and when thermalization occurs, the relic density plateaus.  This is because the Freeze-In mechanism has become the Freeze-out scenario where the mediator mass no longer affects the relic density. Thermalization is reached in these regions of the parameter space as $g_V^2g_D^2 \sim m^4_{X}\sigma v/ \mDM^2$ and hence for a fixed cross-section and DM mass, both the visible and dark sector couplings grow as $m_X$ is increased.  While we have chosen a point in the model parameter space that does not suffer from the hot dark matter constraint, we find that for sufficiently large $m_{X}\gtrsim 10^{2.5}\,\text{GeV}$ mediator, thermalization tends to occur. The effect of choosing a smaller annihilation cross-section would mean that thermalization occurs for larger mediator masses, however as we observed from the previous section, smaller cross-sections tend to be more vulnerable to the warm DM constraint.

\subsection{The effect of PBH spin}
\begin{figure}[t!]
\centering
 \includegraphics[width=\linewidth]{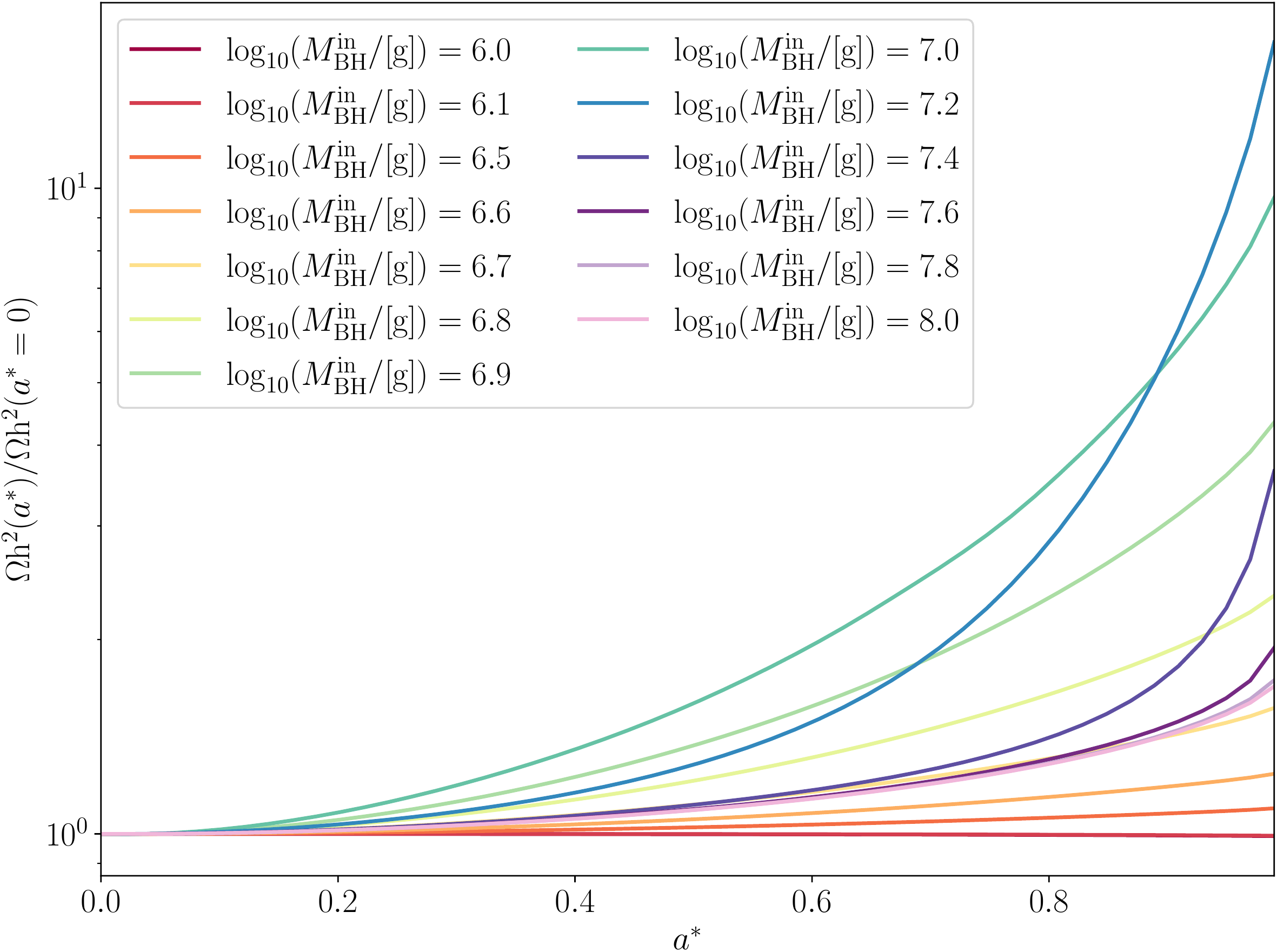}
 \caption{\label{fig:FI_spin} The  relic density as a function of the PBH spin, $a^*$, for the FI mechanism with $\beta^\prime = 10^{-10}$, 
 $\mDM = 10^{-3.0}\, \rm{GeV}$  and $\mathrm{Br}(X\to \mathrm{DM}) = 0.5$ and $\log_{10}(\sigma v/[\rm{GeV}^{-2}]) = -42.0$ for
initial PBH mass of  various initial PBH masses.}
 \end{figure}
 
 In the previous section, we considered the effect of Schwarzschild PBHs on the FI mechanism and identified regions where DM remains cold and does not thermalize: $10^{-3}\lesssim \mDM\,(\text{GeV}) \lesssim 10^{-2}$
 and $10\lesssim m_{X}\,(\text{GeV})\lesssim 10^{1.5}$.
 However, Hawking radiation rates are sensitive to the angular momentum of the PBH \cite{Page:1976ki}, and in this Section, we briefly summarize the effect of PBH spin on the FI mechanism.  A more detailed discussion of the effect of spin on Hawking evaporation can be found in our companion paper \cite{paper1}.
 
 In \figref{fig:FI_spin}, we show how the relic density varies as a function of the PBH spin, $a^*$, for various initial PBH masses with $\beta^\prime = 10^{-10}$, 
 $\mDM = 10^{-3.0}\, \rm{GeV}$  and $\mathrm{Br}(X\to \mathrm{DM}) = 0.5$ and $\log_{10}(\sigma v/[\rm{GeV}^{-2}]) = -42.0$. With all parameter fixed, apart from the spin of the PBH, the 
relic density increases as a function of $a^*$ for $\log_{10}(\MBH^{\rm in}/[\rm{g}])\gtrsim 10^{6.1}$\footnote{For smaller PBH masses, PBH domination does not occur and therefore varying the spin will not affect the relic density.}. This is because spinning PBHs evaporate faster than non-spinning PBHs and if the evaporation temperature is close to the mass of the mediator, then the thermal plasma can be reheated when Freeze-In is most effective and this contributes to an enhancement of the relic density. For instance, 
the largest enhancement, for this given point,  occurs for $\log_{10}(\MBH^{\rm in}/[\rm{g}])= 7.2$ where the effect of spin can enhance the relic density by a factor $\sim 15$. For this point in the parameter space, the evaporation temperature without spin is  $m_{X}/ T_{\rm ev, a^*=0} \sim 5$, while with spin,  $m_{X}/T_{\rm ev, a^*=0.99} \sim 3$. For larger initial PBH masses, $\log_{10}(\MBH^{\rm in}/[\rm{g}]) > 7.2$, the evaporation temperature is much lower than the mediator mass and thus the presence of spinning PBHs still increase the relic abundance, relative to non-spinning PBHs, but the effect is not as significant. 

Interestingly this implies that the dilution effect discussed in \secref{sec:FI}, can be mitigated if 
the PBHs have a significant spin.

\section{Discussion}
The evaporation of primordial black holes constitutes a natural source of gravitational production for any dark sector of particle physics. This suggests that, although the FO or FI mechanisms can lead the SM bath to produce DM particles thermally, they might not be the only contribution to the final DM relic abundance.

Two main effects of this co-production of DM from PBH evaporation and thermal processes have been studied in the literature. First, the extra contribution of PBHs to the final relic abundance leads to readjust the amount of DM particles expected from the usual thermal production to match with observations. Second, it was known that PBHs could transiently dominate the energy density of the Universe. Their subsequent evaporation into visible entropy can lead, in that case, to a dilution of a preexisting DM relic abundance. However, this evaporation was always treated as instantaneous in the computation of the dilution in previous works. In this paper, we addressed a few caveats inherent to such an approach by studying a vanilla model including a fermionic DM, $\psi$, which couples to SM particles through the exchange of a vector mediator, $X$. In particular, we focused on the case where the mediator is heavier than DM but has a mass smaller than the reheating temperature. 

In the case of the Freeze-Out mechanism, we studied three different scenarios:
when PBHs evaporate before FO; we checked that the DM particles produced from evaporation thermalize with the SM and therefore do not affect the dynamics of the FO mechanism nor contribute to the final relic abundance. In the case where the evaporation takes place after the FO, there exist two possible situations: 
\begin{itemize}
    \item[(i)]  PBHs may dominate the energy density of the Universe before evaporating. If the FO takes place before or during a PBH dominated era (which can typically take place for $\mDM\lesssim 1\mathrm{GeV}$), we derived useful analytical results (based on the findings of Ref.~\cite{Arias:2019uol}) in order to calculate the relic density produced through FO and showed that the evaporation of PBHs can affect the prediction of the standard FO scenario during PBH domination. However, we showed that it is not possible that PBHs can both affect the FO production dynamics and contribute significantly to the relic density at the same time if they evaporate after the Freeze-Out time unless they destroy the FO predictions by interacting efficiently with the thermal DM remnants. This is due to the mediator's presence, which both provides additional annihilation channels for DM and leads to resonant interactions between DM and SM particles. Indeed, when they are produced through evaporation, DM particles carry a momentum $p_{\rm DM}\sim T_{\rm BH}\gg m_X$, which after redshifting long enough can hit the resonance at $p_{\rm DM}\sim m_X$. Therefore, we find that the interaction of evaporated DM particles with the preexisting relic abundance or with SM particles have to be efficient if PBHs dominate the universe energy density before they evaporate. In that case, the non-trivial phase-space distribution of the evaporation products requires solving the Boltzmann equation at the level of phase-space distributions, which we let for future work.
\item[(ii)] In the case where PBHs never dominate the energy density, the contributions of the FO mechanism and PBH evaporation may sum up without affecting each other. 
 \end{itemize}

In the FI case, the particle interactions are small enough to avoid the thermalization constraint easily. When the temperature reaches $T\sim m_X$, most of the FI production takes place. In the case where PBHs dominate the energy density of the Universe before evaporating, the effect of PBH evaporation on the FI production all depends on whether this happens before, during or after PBH evaporation. We went beyond the instantaneous evaporation approximation and derived analytical results for the FI production. We scanned over the parameter space and showed that the effect of the PBH evaporation could lead to requiring an annihilation cross-section which is larger by several orders of magnitude than in the standard FI case. Thanks to the results of Ref.~\cite{paper1}, we also derived the precise averaged momentum of the evaporated DM particles and constrained the model using Lyman-alpha constraints on warm DM. 

We studied the influence of the mediator mass, as well as the possible BH spin on the value of the relic density in the Freeze-In scenario. We showed that those additional degrees can play an important role in the choice of the cross-section which is necessary in order to obtain the observed relic abundance.

Finally, we comment on the general effect of PBH evaporation on the possible detection of DM in the future. In the FO case our findings suggest that the main effect of the evaporation of PBHs on the thermal production can take place when PBHs dominate the energy density and evaporate during and after the FO. As we have seen, the evaporation products are expected to scatter efficiently with the FO relic density and with the SM bath. Our guess is that the main effect of this evaporation would therefore be to wash-out the relic density produced by FO. Although this would have to be confirmed by an appropriate solving of the Boltzmann equation, this could allow the FO mechanism to take place with smaller values of the annihilation cross-section and therefore tend to escape detection.

In the case of the FI, we have thoroughly studied how the entropy injection during and after FI can lead to significantly larger cross-sections in order to obtain the correct relic abundance. This effect enhances by orders of magnitude the detectability of our FI scenario in certain regions of the parameter space.

\section*{Acknowledgment}
The authors would like to thank Mathias Pierre for useful discussions. LH would like to thank Shuo Pan for particularly supportive conversations during the realization of this work. The work of LH is funded by the UK Science and Technology Facilities Council (STFC) under grant ST/P001246/1. This manuscript has been authored by Fermi Research Alliance, LLC under Contract No. DE-AC02-07CH11359 with the U.S. Department of Energy, Office of Science, Office of High Energy Physics.


\appendix

\section{Decay Widths and Cross-Sections}\label{app:Xsections}
The mediator $X_\mu$ is unstable on cosmological scales and its decay width is given by
\be
\Gamma_X = \Gamma_{X\to \mathrm{DM}}+\Gamma_{X\to \mathrm{SM}}\,,
\ee
where
\bea \label{eq:GamXDM}
\Gamma_{X\to \mathrm{DM}}&=&\frac{\gD^2 }{12\pi}M_X\left(1+\frac{2\mDM^2}{M_X^2}\right)\sqrt{1-\frac{4 \mDM^2}{M_X^2}}\,,\nonumber\\
\Gamma_{X\to \mathrm{SM}}&=&\frac{\gV^2 }{12\pi}M_X\left(1+\frac{2 m_f^2}{M_X^2}\right)\sqrt{1-\frac{4 m_f^2}{M_X^2}}\,.
\eea

In this case, one obtains
\be
\int\dd \Omega |\overline{\mathcal M}|^2 = 12\pi g_{\rm SM}^2g_\chi^2\frac{(s+2m_\chi^2)(s+2 m_{\rm SM}^2)}{(s-M_X^2)^2+M_X^2 \Gamma_X^2}\,,
\ee
which gives, using Eq.~\eqref{eq:sigmav},
\be
\ba
&\langle\sigma v\rangle = \frac{\pi g_{\rm SM}^2g_\chi^2 T^4}{3(2\pi)^6 n_{\chi,\mathrm{eq}}^2} \\
& \int_{2 x}^\infty (z^2-4 x^2)^{\frac{1}{2}} z\frac{(z^2T^2+2 m_\chi^2)(z^2T^2+2 m_{\rm SM}^2)}{(z^2 T^2 - M_X^2)^2+M_X^2 \Gamma_X^2} K_1\left[z\right] \dd z\,.
\ea
\ee
In the narrow width approximation, this expression becomes
\be
\ba
&\langle\sigma v\rangle = \frac{3 g_D^2g_V^2 T}{512 \pi^5 n_{\chi,\mathrm{eq}}^2} \\
&(M_X^2-4 m_\chi^2)^{\frac{1}{2}}\frac{(M_X^2+2 m_\chi^2)(M_X^2+2 m_{\rm SM}^2)}{M_X \Gamma_X} K_1\left[\frac{M_X}{T}\right]\,.
\ea
\ee
\section{Thermally-Averaged Cross-Sections with Different Temperatures}\label{app:XsectionsdiffT}
Significant thermalization of DM particles produced by PBH evaporation is important in this work
and indicates when our momentum-averaged Boltzmann equations are not reliable. To determine if
thermalization has occurred, we calculate the interaction rate of the PBH produced DM particles with the
DM particles produced from the thermal bath. As the temperatures of these two populations
can differ, we must go beyond the standard thermal-averaging calculation \cite{Gondolo:1990dk}. In this Appendix, we present the 
derivation of the thermally-averaged cross-section of two particles with differing temperatures:
\be\label{eq:cx}
\langle\sigma \cdot v\rangle_{T_{1} T_{2}}=\frac{\int \sigma \cdot v f_{1} f_{2} d^{3} \vec{p}_{1} d^{3} \vec{p}_{2}}{\left[\int d^{3} \vec{p}_{1} f_{1}\right]\left[\int d^{3} \vec{p}_{2} f_{2}\right]}
\ee
where the temperature and phase-space distribution of $\psi_1$ ($\psi_2$) is given by 
 $T_1$ ($T_2$) and $f_1$ ($f_2$) respectively. We approximate the phase-space distribution of the particles to be Maxwellian i.e. $f\propto \exp\{-E/T\}$
 The phase-space terms can be written as a function of particle energies, three momenta and angular separation between the three momenta ($\theta$):
 \be
d^{3} \vec{p}_{1} d^{3} \vec{p}_{2}= 16 \pi^2 \vec{p}_{1} E_{1} d E_{1} \vec{p}_{2} E_{2} d E_{2} d \cos \theta\,.
\ee
Applying a change of variables:
 \be
 \ba
x_{+} &\equiv \frac{E_{1}}{T_{1}}+\frac{E_{2}}{T_{2}}\,,\\
x_{-} & \equiv \frac{E_{1}}{T_{1}}-\frac{E_{2}}{T_{2}}\,,\\
s & \equiv 2 m^{2}+2 E_{1} E_{2}-2 \vec{p}_{1} \vec{p}_{2} \cos \theta\,,
 \ea
\ee
assuming $\psi_1$ and $\psi_2$ have the same mass, $m$, the phase-space can be written as
\be
d^{3} p_{1} d^{3} p_{2}=2 \pi^{2} T_{1} T_{2} E_{1} E_{2} d x_{-}d x_{+}d s\,.
\ee
Therefore, the numerator of Eq.~(\ref{eq:cx}) can be written as
\be
2 \pi^{2} T_{1} T_{2} \int \sigma \sqrt{(p_1\cdot p_2)^2-m^4 } \,e^{-x_{+}} \,d x_-d x_+d s\,.
\ee
The integration region ($E_1, E_2 >m$ and $\lvert \cos\theta \rvert \leq 1$) we find that this numerator is
\be\label{eq:cx1}
4 \pi^{2} T_{1} T_{2} \int_{4 m^{2}}^{\infty} d s \,\sigma \, \sqrt{(p_1\cdot p_2)^2-m^4 } \int_{x^{\min}_{+}}^{\infty} d x_+x_{-}^{\max } e^{-x_{+}}\,,
\ee
where 
\be
\ba
x_{+}^{\min} & = \frac{\sqrt{m^{2}\left(T_{1}-T_{2}\right)^{2}+ s T_{1} T_{2}}}{T_{1} T_{2}} \,,\\
x_{-}^{\max} & = \frac{x_+ A +\left[s(s-4m^2)(T_1^2T_2^2x^2_{+}-B)\right]^{\frac{1}{2}}}{B}
\ea
\ee
with
\be
\ba
A & = m^{2}\left(T_{2}^{2}-T_{1}^{2}\right)\,,\\
B& = m^{2}\left(T_{1}-T_{2}\right)^{2}+ s T_{1} T_{2}\,.
\ea
\ee
The integration over $x_+$ yields
\be\label{eq:cx2}
\ba
\int_{x_{+}^{\min }}^{\infty} d x_+e^{-x_{+}} x_{-}^{\max }= & 
=&\frac{A}{B}\left\{1+z\right\} e^{-z}+\frac{C}{\sqrt{B}} K_{1}\left(z \right)\,.
\ea
\ee
where 
\be
\ba
z &= \sqrt{m^{2}\left(\frac{T_{1}-T_{2}}{T_{1} T_{2}}\right)^{2}+\frac{s}{T_{2} T_{2}}}\,,\\
C &= \left(s\left(s-4 m^{2}\right)\right)^{\frac{1}{2}}\,.
\ea
\ee
Substituting \equaref{eq:cx2} into \equaref{eq:cx1}, we find that the numerator of \equaref{eq:cx} is
\bea\label{eq:num}
&&2 \pi^{2} T_1 T_{2} \int_{4 m^{2}}^{\infty} ds \sigma \frac{\sqrt{s(s-4m^2)}}{B}\nonumber\\
&\times&\left( A\left\{1+z\right\} e^{-z}+C\sqrt{B} K_{1}\left(z\right)\right)\,.
\eea
\begin{widetext}
We note that in the limit $T_1=T_2$, the numerator expression above simplifies to that found in 
\cite{Gondolo:1990dk} as expected.
The denominator of \equaref{eq:cx} is more straightforward and is given by 
\be\label{eq:den}
16 \pi^{2} m^{4} T_{1} T_{2} K_{2}\left(\frac{m}{T_{1}}\right) K_{2}\left(\frac{m}{T_{2}}\right)
\ee
Combining \equaref{eq:num} and \equaref{eq:den}, we find that the thermally-averaged cross-section is 
given by
\bea\label{eq:den}
&&\langle\sigma \cdot v\rangle_{T_{1} T_{2}}= D \int_{4 m^{2}}^{\infty} d s\, \sigma \frac{C}{B}\left(A\left(1+z\right) e^{-z}+C \sqrt{B} K_{1}\left(z\right)\right)
\eea
where
\be
D=\frac{1}{8 m^{4} K_{2}\left(\frac{m}{T_1}\right){K_{2}}\left(\frac{m}{T_{2}}\right)}.
\ee
Finally, performing a simple transformation to integrate with respect to $z$ instead of $s$, we obtain
\begin{align}
\langle\sigma \cdot v\rangle_{T_{1} T_{2}}= D \int_{z_{\rm min}}^{\infty} d z \,\sigma C \left(\frac{A}{T_ 1T_2}\frac{\left(1+z\right)}{z} e^{-z}+C K_{1}\left(z\right)\right),  
\end{align}
being $z_{\rm min}=m(T_1+T_2)/(T_1 T_2)$.
\end{widetext}

\bibliography{main}
\end{document}